\documentclass[aps,prfluids,reprint,groupedaddress,superscriptaddress,onecolumn,a4paper,amsfonts,amssymb,amsmath]{revtex4-2}

\usepackage{hyperref}
\usepackage{graphicx}
\usepackage{amsmath}
\usepackage{amssymb}
\usepackage{bm}
\usepackage{units}
\usepackage{upgreek}
\usepackage{float}
\usepackage{cleveref}
\usepackage{cancel}
\usepackage{tikz}

\usepackage{array}

\usepackage{xcolor}
\hypersetup{
    colorlinks,
    linkcolor={red!50!black},
    citecolor={blue!50!black},
    urlcolor={blue!80!black}
}

\definecolor{FU}{RGB}{0, 51, 102}
\definecolor{FUlight}{RGB}{0, 102, 204}
\definecolor{FUred}{RGB}{204, 0, 0}
\definecolor{FUgreen}{RGB}{153, 204, 0}
\crefname{equation}{Eq.}{Eqs.}
\Crefname{equation}{Eq.}{Eqs.}

\crefname{figure}{Fig.}{Figs.}
\Crefname{figure}{Fig.}{Figs.}

\crefname{table}{Tab.}{Tabs.}
\Crefname{table}{Tab.}{Tabs.}

\definecolor{Turk}{rgb}{0,0.7,0.4}

\newcommand{\tsed}{\tau_\mathrm{sed}}
\newcommand{\tev}{\tau_\mathrm{ev}}
\newcommand{\kb}{k_\mathrm{B}}

\def\R{\mathbb{R}}
\newcommand{\bs}[1]{\boldsymbol{#1}}

\begin{document}

\title{
Combined effects of evaporation, sedimentation and  solute crystallization on the 
dynamics of aerosol size distributions on multiple length and time scales
}

\author{Sina Zendehroud}
\affiliation{Freie Universit\"at Berlin, Department of Physics, Arnimallee 14, 14195 Berlin, Germany}

\author{Ole Kleinjung}
\affiliation{Freie Universit\"at Berlin, Department of Physics, Arnimallee 14, 14195 Berlin, Germany}

\author{Philip Loche}
\affiliation{Laboratory of Computational Science and Modeling, Institute of Materials, \'Ecole Polytechnique F\'ed\'erale de Lausanne, 1015 Lausanne, Switzerland}

\author{Lyd\'{e}ric Bocquet}
\affiliation{Université PSL, Laboratoire de Physique de l'École Normale Supérieure, Paris 75005, France}

\author{Roland R. Netz}
\affiliation{Freie Universit\"at Berlin, Department of Physics, Arnimallee 14, 14195 Berlin, Germany}

\author{Erica Ipocoana}
\affiliation{Freie Universit\"at Berlin, Department of Mathematics and Computer Science, Arnimallee 9, 14195  Berlin, Germany}

\author{Dirk Peschka}
\affiliation{Weierstrass Institute for Applied Analysis and Stochastics, Wilhelm-Anton-Amo Str.\ 39, 10117 Berlin, Germany}

\author{Marita Thomas}
\affiliation{Freie Universit\"at Berlin, Department of Mathematics and Computer Science, Arnimallee 9, 14195  Berlin, Germany}

\date{\today}

\begin{abstract}
We investigate three aspects of 
aerosols produced by humans speaking, coughing or sneezing on vastly different length and time scales, which are potentially important for understanding
aerosol-mediated air-borne viral infection mechanisms.
First, we address  the evolution of the size distribution of a non-interacting ensemble of droplets that are subject
to  evaporation and sedimentation  using  a  sharp droplet-air interface model. 
From the exact solution of the 
evolution equation governing the aerosol size distribution, we derive the total time-dependent fraction of virions suspended in the air and show that it decreases with increasing relative humidity of the ambient air. This not only explains experimental findings but also indicates a practical method to reduce the airborne viral  infection risk.
 Secondly, from Molecular Dynamics simulations we extract the molecular reflection coefficient
 of single water molecules that encounter the vapor-liquid water interface from the vapor phase. The molecular reflection coefficient determines the water-vapor condensation rate   at a liquid droplet surface and also the evaporation rate from aqueous droplets. We find
 the reflection of water to be negligible at room temperature but to rise significantly at elevated temperatures
 and for grazing incidence angles. 
Thirdly, we derive a thermodynamically consistent three-dimensional diffuse-interface model for  
solute-containing droplets 
that  is  formulated as a three-phase Cahn-Hilliard/Allen-Cahn system.
By numerically solving the coupled system of equations, 
we explore representative scenarios and show that this model reproduces and generalizes features of the sharp-interface model.
These interconnected studies on  the dynamics of aerosol droplet evaporation might in the future help  to understand the airborne infection risk under varying environmental conditions. 
\end{abstract}

\maketitle

\section{Introduction}
\label{sec:intro}
Speaking, coughing, and sneezing  produce aerosols of saliva droplets
\cite{Johnson2011,Bourouiba2014,Duguid1946,Bozic2021}, which, depending on their size, fall to the ground quickly
or evaporate and remain suspended in the air for extended times. Accordingly, such droplets containing viruses (or more precisely virions, where we will use both terms interchangeably in this paper) from infected persons, which remain suspended in the air for prolonged times, make the environment hazardous. Aerosols
are known to be vectors of virus spreading, as shown convincingly for influenza
\cite{Cowling2013,Shaman2009,Shaman2010,Tellier2009,Lowen2009}. For SARS-CoV-2,
the results of available studies are consistent with virus aerosolization 
from normal breathing, following several reports
indicating that viruses can float in aerosol droplets for hours and
remain infectious \cite{Doremalen2020}, together with evidence for broad dispersion
of RNA in  isolation rooms, which
indicates that viruses can spread via aerosols \cite{Santarpia2020}.

For many practical biomedical questions a key factor is the evaporation-driven evolution of aerosol droplet size and resulting particle morphology, which depends on the concentration and type of solute in the droplets.
Virus-laden aerosols typically have a complex composition and contain various solutes and macromolecules, which can induce  complex internal fluid flows and compositional heterogeneities.  Here, we restrict ourselves
to one-component low-molecular weight solutes.
 Solutes can significantly alter evaporation rates and the morphology of the particles formed when aerosol droplets dry, see \emph{e.g.} \cite{LEONG1987511}. 
Typical particle morphologies produced upon evaporation are hollow spheres, porous spheroids or solid spheres and the particles can be single crystals, polycrystalline or weakly bound agglomerates.
These properties of dried particles depend on various parameters such as the evaporation rate, the initial solute concentration, the ambient humidity and temperature.

Previous studies of droplet drying, crust formation, and crystallization have primarily employed one-dimensional or spherically symmetric moving-boundary models and population balance approaches to describe solute transport, nucleation, and shell formation \cite{LEONG1987511, gregson2019drying, rezaei2021water, vehring2007particle}. 
Experimental studies of single droplets provide detailed measurements of drying and crystallization kinetics and, together with theoretical modelling, allow one to infer internal solute concentration profiles \cite{gregson2019drying}. The onset of crust formation and its impact on drying have also been studied theoretically \cite{rezaei2021water,Rezaei2021}, while recent experimental and modeling studies have provided detailed measurements of evaporation and crystallization kinetics in levitated saline droplets \cite{Mignot-2026}. These processes are relevant not only for aerosol particles, but also for spray-drying applications \cite{vehring2007particle}.  
The coupled transport and phase transition mechanisms that govern single-droplet evaporation have been comprehensively reviewed \cite{sadek-2015}, where the authors specifically detail how internal fluid dynamics and solute diffusion rates drive the process of surface shell formation and crystallization, which ultimately dictate the final morphology of the dried particle. However, the above approaches are best suited to isolated droplets and become difficult to extend to aerosol systems where symmetry is broken by spatial heterogeneities, external vapor gradients, or droplet coalescence. In such cases, local variations in the vapor chemical potential and crystallization dynamics can influence the resulting shell morphology.

Phase-field methods provide an alternative framework that avoids explicit interface tracking. 
Diffuse-interface models based on the Cahn-Hilliard and Allen-Cahn equations are widely used to describe interfacial dynamics in multiphase systems. The coupling to the Navier-Stokes equation has been studied for two-phase and multiphase flows \cite{ADG-2013,Eikelder-2025,boyerlapuerta-2006,ggps-ch}, including formulations with soluble surfactants \cite{Voigt-2011}. Related phase-field approaches have also been used to describe crystal growth and nucleation processes \cite{Voigt-2009}, phase separation processes with interfacial chemical reactions in ternary mixtures with surfactants \cite{ToZhYA02PSDR}  and the dynamics of droplets on deformable membranes \cite{Aland-2024}. 
Applications include  evaporating crystalline multicomponent films \cite{Ronsin2022}, multicomponent droplet evaporation \cite{Yang-2022}, solute precipitation and dissolution \cite{xu2008phase, Lamperti2026}, evaporation in porous media \cite{Ghosh-2025}, vapor-liquid-solid crystallization during nanowire growth \cite{Wang-2018}, phase separation in polymer-polymer-solvent mixtures undergoing  evaporation of the solvent phase \cite{LyCiMu24PSMF}, as well as  the evaporation and crystallization of saline droplets \cite{Mignot-2026}. 
Motivated by these observations, our aim is to  investigate droplet evaporation dynamics in connection to airborne infection risk.
In Section II, we study the evolution of the size distribution of droplets due to evaporation and sedimentation, and present an exact solution of the  equation that governs the dynamics of the aerosol size distribution. 
Here we employ the simplifying sharp droplet-air interface model, which allows us to quantify the effect of humidity on the droplet size distribution dynamics in closed form.
Our results show that increasing humidity drastically reduces evaporation and thereby reduces the 
sedimentation time of aerosols, which in turn decreases the 
number of viruses that remain airborne in case the aerosol particles are produced by an infected human.
This is in accord with experimental findings and also indicates a simple  method of how to reduce the airborne viral  infection risk.

An important but rather elusive  parameter in the modeling of condensation-evaporation equilibria at vapor-liquid interfaces is the molecular reflection coefficient, which quantifies the fraction of molecules that are reflected from the vapor-liquid interface as a molecule encounters the interface from the vapor phase
and which determines the interfacial molecular adsorption kinetics.
The fraction of molecules that are not reflected, which in literature is denoted as the condensation  or mass-accommodation coefficient \cite{Tobias2004}, stays adsorbed to the interface and therefore condenses onto the liquid phase. 
Experimentally, the molecular reflection coefficient has been extensively discussed, but its value could not be unambiguously determined, see the discussion in  \cite{Tobias2004}.
Molecular Dynamics simulations determined the molecular reflection coefficient of water from a liquid water interface,  but the velocity and incident angle dependencies were not clearly resolved 
\cite{Morita2004,Tsuruta2004,Tobias2004}.
In Section III we use Molecular
Dynamics simulations to estimate the molecular interfacial water reflection coefficient for single water molecules from the vapor-liquid water interface
and show that it depends on the
impinging angle of water molecules as well as on the impinging
velocity. In particular, we show that the reflection coefficient is practically zero for a vapor temperature close to room temperature, such that molecular water reflection can be safely neglected when predicting the evaporation-condensation  kinetics using diffusion-reaction equations in Sections II and IV.

In Section IV  we introduce a three-dimensional, diffuse-interface model formulated as a three-phase Cahn-Hilliard/Allen-Cahn system \cite{ch, EL}, featuring a liquid, a vapor, and a crystalline phase, to generalize the   one-dimensional sharp-interface model discussed in Section II. This three-phase  model  is coupled with a diffusion equation describing the  dynamics of the  solute concentration inside the droplet. The solute could correspond to salt that is initially dissolved in the droplet. With our model we are therefore able to study the dynamic process of salt crystallization and precipitation, for which a diffuse-interface model is preferred over a sharp-interface model, because in the latter the emergence of a new boundary is much harder to implement.
More generally speaking, the solute could also correspond to other biologically relevant constituents, such as dispersed viruses or soluble proteins. In particular, through numerical calculations  we verify that our diffuse-interface model generalizes and is able to reproduce key features of our sharp-interface model studied in Section II. In fact, the generalization is two-fold: Firstly, we treat three phases including an additional solute constituent. Secondly, thanks to the phase-field approach, we can in principle treat non-spherically-symmetric droplet geometries and even changes in the droplet topology.


The paper is organized as follows. The dynamics of the droplet size distribution is discussed in \Cref{sec-sed-model}, where in \Cref{sec:population}, we extend the theoretical framework developed in
Refs.~\cite{Netz2020a,Netz2020} and describe the evolution of an initial distribution of droplets due to evaporation and sedimentation. In \Cref{sec:reflection}, we use Molecular Dynamics simulations to
estimate the molecular interfacial water reflection coefficient, which quantifies the effect of imperfect accommodation of water molecules at the droplet surface as they encounter the surface from the vapor phase and is needed in the evaporation model formulated in Section II. In Section \ref{sec-CH} the radial sharp-interface model for the evaporation of a single liquid droplet, used in Section II, is generalized by means of a diffuse-interface model to describe evaporation and crystallization dynamics in droplets containing solutes. 
The diffuse-interface model for evaporation and crystallization of solute-containing droplets is derived in its general form in \Cref{derivation}, for which we give a weak formulation in \Cref{WeakF}. After addressing the choice of free energy in \Cref{DiscussFreeEnDiss}, we proceed to discretize the system and present and discuss meaningful numerical examples in \Cref{NumEx}. Moreover, the evolution  of multiple droplets with non-symmetric geometry and topology changes is illustrated in Section \ref{sec-conclusion}.
\par
The different approaches discussed in this work provide complementary tools to  describe the evaporation dynamics of single droplets and  droplet ensembles under varying environmental conditions on multiple length and time scales: 
While phase-field models in Section IV allow one to model diffusive processes, phase change, and interaction of a small number of droplets in spatially resolved detail, the population dynamics model in Section II describes the evolution of an ensemble of evaporating, sedimenting, non-interacting droplets using the sharp-interface model. The Molecular Dynamics simulations in Section III describe the condensation of single water molecules in atomistic detail and thereby provide the reflection coefficient, which is a parameter that appears in the sharp-interface model in Section II, but, due to the enormous numerical efforts and limited computer power of Molecular Dynamics simulations,  can only treat the short-time dynamics of  small interfacial systems.

\section{Evolution of droplet size distribution in the presence of evaporation and sedimentation}
\label{sec-sed-model}
In terms
of length and time scales, the typical size of aerosols produced by speaking  is in the micron range \cite{Johnson2011,Bourouiba2014,Duguid1946,Yang2011}, while the
corresponding sedimentation and evaporation times span several orders of magnitude
from milliseconds to hours, depending on the size of the aerosol particles. 
The theoretical description of the evaporation and sedimentation  of liquid droplets as a function of the humidity of the ambient atmosphere was worked out in a number of foundational papers \cite{Wells1934,Duguid1946,Ranz1952} and was  reviewed in
\cite{Bozic2021}.
Droplets evolve due to two main mechanisms: evaporation and
sedimentation \cite{Netz2020a,Netz2020}. The small ones evaporate quickly but remain suspended for a
very long time  in the air: In the absence of solutes, a droplet with initial radius $\unit[5]{\upmu m}$
evaporates in $\unit[120]{ms}$ (for a relative humidity of 50\,\%), a much shorter time than it takes the droplet  to sediment to the ground. In contrast,
a droplet of initial radius
$\unit[55]{\upmu m}$ evaporates in 14.5 seconds but  takes only  7.6 seconds to fall
to the ground from an initial height of 2 meters \cite{Netz2020a}. If a droplet evaporates before touching the ground, hence
reaching a completely dried-out state, it remains in the air for very long times (hours to days), since
Brownian motion counteracts gravity for submicron particles. 
More recently, the evaporation of droplets that contain variable initial concentrations of solutes including the influence of the evaporation cooling on the evaporation speed was addressed. It was found that the solute osmotic pressure as well as the evaporation cooling slow down evaporation and therefore decrease the droplet sedimentation time~\cite{Netz2020a,Netz2020}.
In this section, we consider the problem of aerosol evaporation and sedimentation of many droplets that are described by a time-dependent aerosol size  distribution and calculate 
 the total number of viruses which remain suspended in
the air for a given time. We address also the effect of humidity on the aerosol size distribution and find that the
number of viruses floating in the air reduces with increasing humidity. We mention that in the case of Influenza, rising
humidity has indeed been shown to decrease virus transmission
\cite{Shaman2009,Shaman2010,Tellier2009,Lowen2009}, an observation which was
supported by semi-empirical modeling \cite{Yang2011}. We point out that we do not consider
the effect of humidity on the virus infectiousness and only focus on the physical
mechanisms at play. 
An important result
is that the most infectious  droplets, that is, the droplets containing the largest number of viruses,  are not the smaller ones in the initial
distribution of droplets, but rather the larger ones that nevertheless completely evaporate before
reaching the ground. This is so because  the number of viruses in a given droplet is initially given by $N_\mathrm{v}(R) = 4 R_0^3 n_\mathrm{v} \uppi/ 3$, with $n_\mathrm{v}$ the volumetric
density of virus (in saliva) and $R_0$ the initial droplet radius. As an explicit example, the virus number in
 two droplets with initial radii $\unit[1]{\upmu m}$ and
$\unit[100]{\upmu m}$ differs by a factor of $10^6$.
If the  droplet with initial radius $\unit[100]{\upmu m}$ shrinks to a much smaller radius before
touching the ground, it will remain suspended in the air for a long time and contain a large number of virus particles, and hence be extremely infectious  compared to  droplets with much smaller initial radius.

\subsection{Sedimentation and evaporation dynamics of single droplets}
\label{sec:droplet}

In this section, we formulate the basic equations and briefly summarize some main results of Refs.~\cite{Netz2020a,Netz2020}.

\subsubsection{Droplet sedimentation without evaporation}
\label{sec:sedimentation}

The density distribution $p(z, t)$ of droplets at height $z$ and at time $t$  is
given by the diffusion equation \cite{Netz2020a}
\begin{equation}
    \partial_t p(z, t) = D_\mathrm{R} \partial_z^2 p(z, t) + V \partial_z p(z, t)~,
    \label{eq:diffusion}
\end{equation}
where $D_\mathrm{R}$ is the droplet diffusion coefficient and $V$ is the
stationary drift velocity of the droplets, which by balancing the
Stokes friction force that acts on the droplet with the gravitational force is given by
\begin{equation}
    V = \frac{D_\mathrm{R} m g}{\kb T}~,
\end{equation}
with $m$ the mass of the droplet, $g$ the gravitational acceleration,
$k_\mathrm{B}$ the Boltzmann constant, and $T$ the temperature. 
The mean sedimentation time is given by
\begin{equation}
    \tsed = \frac{z_0}{V} = \frac{9 \eta z_0}{2 \rho R^2 g} = \varphi \frac{z_0}{R^2}~,
    \label{eq:tsed}
\end{equation}
where $z_0$ is the initial height of the droplet, the droplet diffusion coefficient is given by the Stokes-Einstein relation
$D_\mathrm{R} = \kb T / (6 \pi \eta R)$, the mass of the droplet  of radius $R$ and mass density
$\rho$ (which is taken to equal the water mass density)
is $m = (4 \pi/ 3) \rho R^3$, 
$\eta$ is the dynamic viscosity of air, and the short-hand notation $\varphi=9 \eta/(2 \rho g)$ is used.

\subsubsection{ Droplet evaporation in the diffusion-limited regime}
\label{sec:stagnant}

The effect of evaporation decreases the droplet radius $R$ during its descent to
the ground, and according to \Cref{eq:tsed} this increases the
sedimentation time.
The evaporative flux of a water droplet is derived in Ref.~\cite{Netz2020a} from the molecular diffusion equation for water vapor, which reads in spherical coordinates as
\begin{equation}
    \partial_t c(r, t) = r^{-2} \partial_r \left( r^2 D_\mathrm{w} \partial_r c(r, t) \right)~,
    \label{eq:diffusion_vapor}
\end{equation}
where $c(r, t)$ is the water vapor concentration at distance $r$ from the center
of the droplet at time $t$, and $D_\mathrm{w}$ is the molecular water diffusion
coefficient in air. The stationary solution of \Cref{eq:diffusion_vapor}, \emph{i.e.}, the solution for $\partial_t c(r, t) = 0$, is given by
\begin{equation}
    c(r) = c_0 (1 + b/r)~,
    \label{eq:stationary}
\end{equation}
where $c_0$ is the ambient water vapor concentration and $b$ is a constant that
remains to be determined. The water flux balance at the droplet surface $r = R$
is given by
\begin{equation}
    J = -4 \uppi R^2 D_\mathrm{w} \frac{\mathrm d}{\mathrm d R} c(R) = 4 \uppi R^2 \left( k_\mathrm{e} c_\mathrm{l} - k_\mathrm{c} c(R) \right)~,
    \label{eq:flux_balance}
\end{equation}
where $k_\mathrm{e}$ and $k_\mathrm{c}$ are the molecular evaporation and
condensation rates, respectively, and $c_\mathrm{l}$ is the water concentration
in the liquid phase. The expression on the left-hand side of
\Cref{eq:flux_balance} describes the diffusive water flux, while the
expression on the right-hand side describes the net flux due to reactive
evaporation and condensation at the droplet surface. Both expressions must be
equal to ensure mass conservation. Using the stationary solution given in
\Cref{eq:stationary}, the constant $b$ can be determined from
\Cref{eq:flux_balance}, which leads to the total water flux
\begin{equation}
    J = 4 \uppi R^2 D_\mathrm{w} \frac{k_\mathrm{e} c_\mathrm{l} - k_\mathrm{c} c_0}{D_\mathrm{w} + k_\mathrm{c} R}~.
    \label{eq:flux}
\end{equation}
In \Cref{eq:flux}, the limit of diffusion-dominated  evaporation is obtained for
$k_\mathrm{c} R \gg D_\mathrm{w}$, which is valid for droplets with radii $R >
\unit[70]{nm}$ \cite{Netz2020a}, while the limit of reaction-dominated evaporation is
obtained for $k_\mathrm{c} R \ll D_\mathrm{w}$. The molecular
condensation rate $k_\mathrm{c}$ is defined as
\begin{equation}
    k_\mathrm{c} = (1 - p_\mathrm{ref}) \bar{k_\mathrm{c}}~,
    \label{eq:kc}
\end{equation}
where $\bar{k_\mathrm{c}} = \sqrt{k_\mathrm{B} T / (2 \pi m_\mathrm{w})}$ is the kinetic
condensation rate with $m_\mathrm{w}$ the mass of a water molecule, and
$p_\mathrm{ref}$ is the molecular reflection coefficient of water molecules that encounter  the droplet surface from the vapor phase \cite{Persad2016}.
Following Ref.~\cite{Netz2020a}, we  assume $p_\mathrm{ref} = 0$ for determining the threshold between the diffusion-dominated and the reaction-dominated regimes,  which is an
approximation that we will demonstrate to be valid for water vapor at room temperature regardless of the incident angle in \Cref{sec:reflection}. From detailed balance it follows $k_\mathrm{e} c_\mathrm{l}= k_\mathrm{c} c_\mathrm{g}$, meaning that for saturated water vapor with concentration $c_\mathrm{g}$ the evaporation and condensation fluxes exactly cancel, which allows us to eliminate the molecular evaporation rate $k_\mathrm{e}$ from the equations.

In the following, we assume that
the evaporation of a droplet at rest occurs in the
diffusion-limited regime, which is valid for radii $R > \unit[70]{nm}$
\cite{Netz2020a}, so that \Cref{eq:flux} can be written as
\begin{equation}
    \frac{\mathrm d}{\mathrm dt} \left( \frac{4 \uppi}{3} R^3(t) \right) = -4 \uppi R(t) D_\mathrm{w} c_\mathrm{g} v_\mathrm{w} (1- RH) = -2 \uppi \theta (1-RH) R(t)~,
    \label{eq:evaporation}
\end{equation}
where $v_\mathrm{w}$ is the
volume of a water molecule in the liquid phase, $RH = c_0 / c_\mathrm{g}$ is the
relative humidity of air, given by the ratio of the ambient water vapor concentration $c_0$ to
the saturated water vapor concentration $c_\mathrm{g}$, and $\theta = 2
D_\mathrm{w} c_\mathrm{g} v_\mathrm{w}$ is a shorthand notation.
\Cref{eq:evaporation} can be solved to give \cite{Netz2020a}
\begin{equation}
    R(t) = R_0 \left( 1 - \frac{\theta (1-RH)}{R_0^2} t \right)^{1/2}~,
    \label{eq:radius}
\end{equation}
where $R_0$ is the initial droplet radius at time $t=0$. The time needed for
evaporation down to a radius at which osmotic effects due to dissolved solutes
within the droplet balance the water vapor chemical potential  can be
approximated as the time needed to reduce the droplet radius to zero, which follows from  \Cref{eq:radius} as 
\begin{equation}
    \tev = \frac{R_0^2}{\theta (1-RH)}~
    \label{eq:tev}
\end{equation}
and defines the evaporation time \cite{Netz2020a}.
Comparing \Cref{eq:tsed} and \Cref{eq:tev}, we  conclude that the combined sedimentation-evaporation process terminates when either the droplet has completely evaporated or when the droplet hits the ground, whichever occurs first. Thus,  the combined sedimentation-evaporation process terminates after a time
\begin{equation}
    \tau^* = \min(\tsed, \tev) = \min \left( \frac{\varphi z_0}{R_0^2}, \frac{R_0^2}{\theta (1-RH)} \right)~,
    \label{eq:taustar}
\end{equation}
which is maximized for a droplet radius of $R^* = (\varphi \theta z_0
(1-RH))^{1/4}$, and the corresponding time scale is given by
\begin{equation}
    \tau^* = \left( \frac{\varphi z_0}{\theta (1-RH)} \right)^{1/2}~.
    \label{eq:taustar2}
\end{equation}
Thus, a droplet with initial radius $R^*$ evaporates completely when it hits the ground, thus, smaller droplets will form so-called droplet nuclei and  stay floating in air for prolonged times.

\subsection{Droplet population size distribution dynamics}
\label{sec:population}

We now consider the dynamics of a  size distribution of droplets that are subject to sedimentation and evaporation, where we neglect the presence of solutes inside droplets such that droplets for which $\tev<\tsed$ holds will stay floating in air forever.
Let the initial distribution of droplet radii be given by $\tilde p_0(R)$. The
total initial volume of the droplets is then
\begin{equation}
    V_0 = \int_0^\infty \mathrm dR \, \tilde p_0(R) \frac{4 \uppi R^3}{3}~.
    \label{eq:V0}
\end{equation}
For the sake of simplicity, we introduce the droplet volume $v$ as a variable
instead of the radius $R$, where the initial droplet volume distribution
$p_0(v)$ is related to the initial radius distribution $\tilde p_0(R)$ through
$p_0(v) \mathrm dv = \tilde p_0(R) \mathrm dR$, and the initial total droplet
volume is given by
\begin{equation}
    V_0 = \int_0^\infty \mathrm dv \, v p_0(v)~.
    \label{eq:V0v}
\end{equation}

The  droplet-distribution dynamics is governed by a balance equation for
$p(v,t)$ given by
\begin{equation}
    \partial_t p(v,t) = -\partial_v \left( \dot v p(v,t) \right) - \frac{1}{\tsed(v)} p(v,t)~,
    \label{eq:population}
\end{equation}
where the terms on the right-hand side account for evaporation and
sedimentation, respectively. The evaporation rate $\dot v = \mathrm d / \mathrm
dt (4 \uppi R^3 / 3)$ is given by \Cref{eq:evaporation}, and the
sedimentation time $\tsed(v)$ is given by \Cref{eq:tsed} with $R = (3 v / 4
\uppi)^{1/3}$. Using the expressions given in \Cref{eq:tsed,eq:evaporation}, \Cref{eq:population} can be rewritten as
\begin{equation}
    \partial_t p(v,t) = 2 \uppi \theta (1 - RH) \left( \frac{3}{4 \uppi} \right)^{1/3} \partial_v \left( v^{1/3} p(v,t) \right) - \left( \frac{3}{4 \uppi} \right)^{2/3} \frac{v^{2/3}}{\varphi z_0} p(v,t)~.
    \label{eq:population2}
\end{equation}

The initial total number of virions in the air is given by
\begin{equation}
    N_\mathrm{v}^0 = n_\mathrm{v} V_0 = n_\mathrm{v} \int_0^\infty \mathrm dv \, v p_0(v)~,
\end{equation}
where $n_\mathrm{v}$ is the initial number of virions per unit volume of droplet fluid.
Ultimately, we want to estimate the number of virions $N_\mathrm{v}(t)$ that
remain in the air at time $t$. Since droplets, which are created with an initial
volume $v_0$ and number of virions $n_\mathrm{v}$, are subject to evaporation,
the density of virions in a droplet increases as the ratio between the initial
and the current droplet volume, such that the total number of virions in a
droplet stays constant. At any time $t$, the total number of virions in the
droplet is thus given by $n_\mathrm{v} v_0$, where $v_0$ is the initial volume
of the droplet, which is in turn a function of the current droplet volume $v$
and time $t$. The airborne droplets, \emph{i.e.}, those that have not completed
sedimentation yet, evolve due to evaporation, and their radius $R$ is related to
the initial radius $R_0$ through \Cref{eq:radius}, which, rewritten in
terms of the droplet volume $v = (4 \uppi / 3) R^3$, defines reversely the
number of virions in a droplet of volume $v$ at time $t$ as
\begin{equation}
    n_\mathrm{v} v_0(v, t) = n_\mathrm{v} v \left( 1 + \left( \frac{4 \uppi}{3 v} \right)^{2/3} \theta (1 - RH)t \right)^{3/2}~.
    \label{eq:virions}
\end{equation}

To estimate the total number of virions that remain airborne at time $t$, one
has to take into account that, due to evaporation, there is a growing population
of droplets with vanishing volume, which remain airborne indefinitely. It is
therefore much simpler to calculate the total number of virions
$N_\mathrm{g}(t)$ that have reached the ground up to time $t$. From the
distribution dynamics given in \Cref{eq:population}, one deduces the rate
of virion deposition to the ground as
\begin{equation}
    \frac{\mathrm d}{\mathrm dt} N_\mathrm{g}(t) = \int_0^\infty \mathrm dv \, n_\mathrm{v} v_0(v,t) \frac{p(v,t)}{\tsed(v)}~,
    \label{eq:deposition}
\end{equation}
where $n_\mathrm{v}$ is the initial virion density and $v_0(v, t)$ is the
initial volume of a droplet with current volume $v$ at time $t$, given by
\Cref{eq:virions}. Note that, due to conservation of virions, the total
number of airborne virions obeys $\dot N_\mathrm{v} = - \dot N_\mathrm{g}$.

\subsubsection{Exact solution for the time-dependent droplet size distribution}
An exact solution of \Cref{eq:population2} can be obtained after a suitable
change of variables. Consider the distribution $g(x, t)$ of squared radii, \emph{i.e.},
$x = R^2$ and $g(x, t) \mathrm dx = p(v, t) \mathrm dv$. Using the definition of
the droplet volume $v = (4 \uppi / 3) x^{3/2}$, one finds $x = (3 v / 4
\uppi)^{2/3}$, and therefore
\begin{equation}
    v^{1/3} p(v, t) = \frac{2}{3} \left( \frac{3}{4 \uppi} \right)^{2/3} g(x, t)~.
\end{equation}
Using this relation, \Cref{eq:population2} simplifies to
\begin{equation}
    \partial_t g(x, t) = \theta (1 - RH) \partial_x g(x, t) - \frac{x}{\varphi z_0} g(x, t)~.
    \label{eq:gx}
\end{equation}
The general solution of \Cref{eq:gx} is  given by
\begin{equation}
    g(x, t) = \exp\left( \frac{x^2}{2 \theta (1 - RH) \varphi z_0} \right) f\left( x + \theta (1 - RH) t \right)~,
\end{equation}
where $f(\cdot)$ is a function to be determined by the initial conditions. Using the
initial condition $g(x, 0) = g_0(x)$, one finds
\begin{equation}
    f(x) = \exp \left( -\frac{x^2}{2 \theta (1 - RH) \varphi z_0} \right) g_0(x)~,
\end{equation}
so that the exact solution for the droplet distribution reads
\begin{equation}
    g(x, t) = \exp \left( -\frac{x t}{\varphi z_0} - \frac{\theta (1 - RH) t^2}{2 \varphi z_0} \right) g_0 \left( x + \theta (1 - RH) t \right)~.
    \label{eq:gxsol}
\end{equation}
There are multiple ways to exploit this exact solution. In the following, we will use it merely to derive the time evolution of the fraction of virions suspended in air.

Following \Cref{eq:deposition}, the rate of virion deposition to the ground
can be written in terms of the squared radius distribution $g(x, t)$ as
\begin{equation}
    \frac{\mathrm d}{\mathrm dt} N_\mathrm{g}(t) = \frac{1}{\varphi z_0} \int_0^\infty \mathrm dx \, n_\mathrm{v} v_0(x, t) x g(x, t)~,
    \label{eq:NgODE}
\end{equation}
where $n_\mathrm{v}$ and $v_0(x, t)$ are given via \Cref{eq:virions}
as
\begin{equation}
    n_\mathrm{v} v_0(x, t) = n_\mathrm{v} \frac{4 \uppi}{3} \left( x + \theta (1 - RH) t \right)^{3/2}~.
\end{equation}
The initial total number of virions is given by the integral over the initial
distribution $g_0(x)$ as
\begin{equation}
    N_\mathrm{v}^0 = n_\mathrm{v} \int_0^\infty \mathrm dx \, \frac{4 \uppi}{3} x^{3/2} g_0(x)~,
\end{equation}
and we define the fraction of sedimented virions at time $t$ as
\begin{equation}
    \phi_\mathrm{g}(t) = \frac{N_\mathrm{g}(t)}{N_\mathrm{v}^0}~.
\end{equation}
Consequently, the fraction of virions still suspended in air at time $t$ is
given by
\begin{equation}
    \phi_\mathrm{s}(t) = 1 - \phi_\mathrm{g}(t) = 1 - \frac{N_\mathrm{g}(t)}{N_\mathrm{v}^0}~.
    \label{eq:suspended}
\end{equation}

\subsubsection{Fraction of suspended virions  as a function of  
    time}
\label{sec:results}

Assuming that the initial droplet radii are distributed normally, the droplet
distribution can be written as
\begin{equation}
    p_0(R) = \frac{1}{\sqrt{2 \uppi \sigma}} \exp \left( -\frac{(R - R_0)^2}{2 \sigma^2} \right)~,
\end{equation}
where $R_0$ is the mean droplet radius and $\sigma$ is the standard deviation of
the distribution. We assume that the corresponding initial squared radius
distribution is then given similarly by
\begin{equation}
    g_0(x) = \frac{1}{\sqrt{2 \uppi \sigma_g}} \exp \left( -\frac{(x - x_0)^2}{2 \sigma_g^2} \right)~,
    \label{eq:gaussian}
\end{equation}
where $x_0 = R_0^2$ and $\sigma_g = 4 R_0^2 \sigma$.

Experimentally, however, droplet distributions produced by speaking or coughing
are best described by log-normal distributions \cite{Johnson2011}, which can be
written as a function of the diameter $D = 2 R$ as
\begin{equation}
    h_0(D) = \frac{\mathrm d p(D)}{\mathrm d \log D} = \frac{C_n}{\sqrt{2 \uppi \sigma} \log \sigma_0} \exp \left( - \frac{(\log D - \log D_0)^2}{2 (\log \sigma_0)^2} \right)~,
\end{equation}
where the values for $D_0$ and $\sigma_0$ are taken from
Ref.~\cite{Johnson2011}, and $C_n$ is a normalization constant, which can be
chosen arbitrarily in our case, since we are only interested in the fraction of
suspended virions. The characteristic droplet radius $R_0$ is then given by $R_0
= D_0 / 2$, and the initial squared radius distribution for the dimensionless
variable $x = (R / R_0)^2$ is given by
\begin{equation}
    g_0(x) = \frac{h_0(D_0 \sqrt{x})}{2 x}~.
    \label{eq:lognormal}
\end{equation}

\begin{figure}
    \centering
\includegraphics{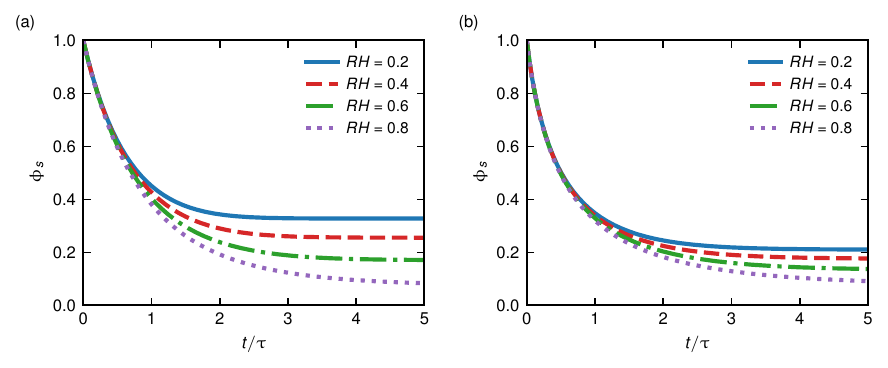}
    \caption{Fraction of suspended virions $\phi_\mathrm{s}(t)$ as a function of rescaled 
    time $t / \tau$ for different relative humidities $RH$, using initial
    droplet distributions given by (a) a normal distribution,
    \Cref{eq:gaussian}, and (b) a log-normal distribution,
    \Cref{eq:lognormal}. The characteristic time scale $\tau$ is given by
the evaporation time for vanishing relative humidity, $\tau = R_0^2 / \theta$,
as follows from  \Cref{eq:tev}.}
\label{fig:suspended_fraction}
\end{figure}

Using the two different  initial distributions in \Cref{eq:gaussian,eq:lognormal}, together with the exact solution for the time evolution,
given in \Cref{eq:gxsol}, we can calculate the fraction of suspended
virions $\phi_\mathrm{s}(t)$, given in \Cref{eq:suspended}, as a function
of time via numerical integration of \Cref{eq:NgODE}.
Following the calculation of $\phi_\mathrm{s}(t)$, one can determine the fraction of
suspended virions in the long-time limit $\phi_\mathrm{s}^\infty =
\lim_{t \to \infty} \phi_\mathrm{s}(t)$ as a function of the relative humidity
$RH$. The results are shown in \Cref{fig:suspended_fraction} for both
initial droplet distributions. We see that for both initial distributions and for all relative humidities $RH$, the fraction of suspended virions $\phi_\mathrm{s}$ decreases considerably after a characteristic time $\tau$, here taken to be the evaporation time for vanishing relative humidity, $\tau = R_0^2 / \theta$,
as follows from  \Cref{eq:tev}.
Up to the time $t\approx\tau$, the effect of $RH$ on the behavior of $\phi_\mathrm{s}$ is minimal. For longer times $t > \tau$, we observe that, for both initial distributions, $\phi_\mathrm{s}$ asymptotically reaches plateau values that depend strongly on both $RH$ and the choice of initial droplet size distribution. We thus conclude that, while the overall behavior of $\phi_\mathrm{s}$ is  qualitatively similar for both normal and log-normal distributions, the choice of initial distribution has a significant effect on the long-time behavior of $\phi_\mathrm{s}$, which becomes more pronounced for low relative humidities.
Obviously, more combined theoretical and experimental research is needed in order to unravel the effects of initial droplet distribution types and relative humidity on the virion number suspended in air.

\section{Molecular interfacial reflection coefficients of water}
\label{sec:reflection}

The molecular reflection coefficient $p_\mathrm{ref}$ of a water molecule that hits the droplet surface from the vapor phase
enters the condensation rate $k_\mathrm{c}$ through \Cref{eq:kc} and
via the detailed-balance condition affects both condensation as well as evaporation rates. While in the diffusion-dominated evaporation regime the resulting evaporation rate becomes independent of the condensation rate and thus also independent of $p_\mathrm{ref}$, as seen in  \Cref{eq:evaporation}, the evaporation rate in the reaction-dominated regime as well as the threshold radius between the two regimes do depend on $p_\mathrm{ref}$.
Generally, the reflection
coefficient $p_\mathrm{ref}$ is  assumed to be close to zero, and in
Ref.~\cite{Netz2020a}, $p_\mathrm{ref} = 0$ was indeed assumed, which corresponds to the limit that each water molecule that hits the droplet surface from the vapor phase actually sticks to the droplet surface.
Experimentally, the molecular reflection coefficient is difficult to measure \cite{Tobias2004}. Previous
Molecular Dynamics simulations determined the molecular reflection coefficient of water from a liquid water interface,  but the temperature and incident angle dependencies were not clearly resolved 
\cite{Morita2004,Tsuruta2004,Tobias2004}.
Using Molecular Dynamics (MD) simulations, we determine the molecular
reflection coefficient $p_\mathrm{ref}$ as a function of the angle and velocity
of single  water molecules that hit a planar vapor-liquid water interface from the vapor phase, where the liquid water is kept at room temperature in the simulations.
This is a highly idealized scenario, but, due to its simplicity, it allows us
to derive reliable estimates for reflection coefficients, which presumably are also relevant for more realistic aerosol scenarios.

\subsection{Simulation setup}
The simulation box used in all simulations has dimensions of $\unit[15]{nm}
\times \unit[5]{nm} \times \unit[5]{nm}$ along the $x$-, $y$-, and $z$-axes,
respectively, and periodic boundary conditions are applied in all three
directions. A block of water measuring $\unit[5]{nm} \times \unit[5]{nm} \times
\unit[5]{nm}$ is placed in the center of the box, such that its edges are
located at $x = \unit[5]{nm}$ and $x = \unit[10]{nm}$, see \Cref{fig:reflection} (a) for a simulation snapshot. The TIP4P water model is employed, which consists of two hydrogen
atoms, one oxygen atom, and an additional massless site representing the
negative charge center. This model was selected because it offers an accurate
representation of hydrogen bonding, the dominant intermolecular interaction
relevant to this work. All molecular dynamics simulations are carried out using the program
GROMACS 2019 with a $\unit[2]{fs}$ integration time step. Temperature coupling
is applied using the velocity rescale thermostat with a stochastic term and a
time constant of $\unit[0.5]{ps}$, maintaining a reference temperature of
$\unit[300]{K}$. The Lennard-Jones interactions are treated using a cutoff
scheme with a cutoff distance of $\unit[1.2]{nm}$ and a potential-shift
modifier. Electrostatic interactions are computed using the particle mesh Ewald
(PME) method with a real-space cutoff of $\unit[1.2]{nm}$ and a Fourier spacing
of $\unit[0.2]{nm}$. All bonds involving hydrogen atoms are constrained using
the LINCS algorithm, and the center-of-mass motion is removed.

\subsection{Vapor phase}
We define the vapor phase as the region more than $\unit[1]{nm}$ away from the
water slab, \emph{i.e.}, $x > \unit[11]{nm}$ and $x < \unit[4]{nm}$. This definition
was chosen to compensate for interfacial fluctuations at the edges of the slab of water. To
validate the simulation setup and ensure stability, we calculate the vapor
pressure of the system, which is obtained via the ideal gas law as
\begin{equation}
    P_\mathrm{vap} = \frac{N_\mathrm{avg} k_\mathrm{B} T}{V}~,
\end{equation}
where $k_\mathrm{B}$ is the Boltzmann constant, $T$ is the temperature, $V$ is
the volume of the vapor phase, and $N_\mathrm{avg}$ is the average number of
water molecules per frame in the vapor phase. The vapor pressure is found to be
$P_\mathrm{vap} = \unit[63.786]{mbar}$,   considerably larger than the experimental value $P_\mathrm{vap}^\mathrm{exp} =
\unit[35.670]{mbar}$ \cite{Lide2004}, which is a well-known property of the TIP4P model. Note that our value for the vapor pressure is in agreement with previous simulations using the TIP4P model \cite{Vega2006}.

\subsection{Reflection simulations}
To determine the molecular reflection coefficient $p_\mathrm{ref}$ at the
vapor-liquid water interface, we perform a series of simulations where a single
water molecule is added to the system outside the liquid water slab
and placed at $\bm{x_0} = (1.783, 2.662, 3.161)^T$. In the following,
the phrase `single molecule' refers to this externally placed molecule. When a
simulation is started, every atom velocity as well as the velocities for the
atoms of the single molecule are randomly assigned according to a
Maxwell-Boltzmann distribution at $\unit[300]{K}$. Consequently, the molecules
also possess center-of-mass velocities that follow a
Maxwell-Boltzmann distribution at $\unit[300]{K}$.
To broaden the range of the analyzed velocities, additional simulations are
performed where the velocity of the single molecule is multiplied by a variable factor of two or four after drawing it from a Maxwell-Boltzmann
distribution at $\unit[300]{K}$. This procedure is equivalent to drawing the
initial velocity of the single water molecule from a Maxwell-Boltzmann distribution at $\unit[1200]{K}$ and
$\unit[4800]{K}$, respectively. For each of these three scenarios, 200,000
simulation runs are performed, leading to a total of 600,000 simulations, each
lasting for $\unit[30]{ps}$.

\begin{figure}[ht]
    \centering
    \includegraphics{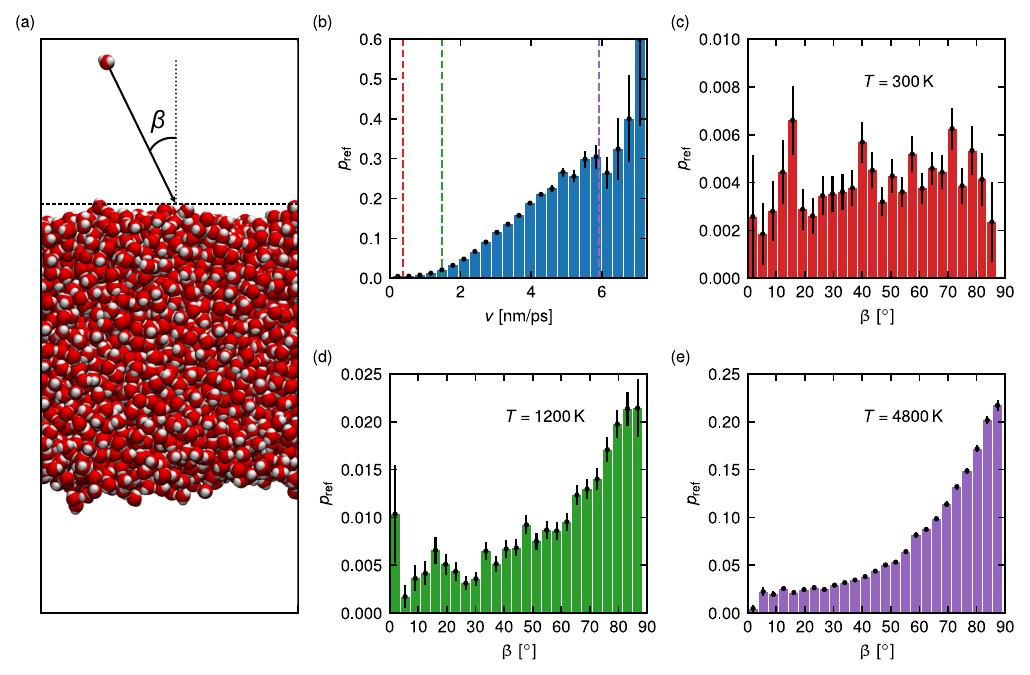}
    \caption{(a) Simulation snapshot. The angle of incidence $\beta$ is defined as the angle between the initial velocity vector of the single water molecule in the vapor phase (black arrow) and the surface normal of the water slab surface (black dotted line).
    (b) Molecular reflection coefficient $p_\mathrm{ref}$ as a function
    of the velocity of the impinging water molecule. The histogram combines data
    from all three sets of simulations with different initial single molecular velocity
    distributions. Average velocities according to $\sqrt{\kb T / m_\mathrm{w}}$
    for $T = \unit[300]{K}$ (red), $\unit[1200]{K}$ (green), and
    $\unit[4800]{K}$ (purple) are indicated by vertical dashed lines. (c)
    Molecular reflection coefficient $p_\mathrm{ref}$ as a function of the angle
    of incidence $\beta$ of the impinging water molecule for initial velocities
    drawn from a Maxwell-Boltzmann distribution at $\unit[300]{K}$. (d) Same as
    (c) but for initial velocities drawn from a Maxwell-Boltzmann distribution
    at $\unit[1200]{K}$. (e) Same as (c) but for initial velocities drawn from a
    Maxwell-Boltzmann distribution at $\unit[4800]{K}$. Error bars in all panels
    indicate the standard error given by \Cref{eq:error}.
    Note that the temperature of the liquid slab is in all simulations $T$=300 K.
    }
    \label{fig:reflection}
\end{figure}

A reflection is defined as follows: if a molecule leaves the vapor phase, \emph{i.e.},
its $x$-coordinate exceeds $\unit[4]{nm}$ while moving in the positive
$x$-direction (for $v_x > \unit[0]{nm/ps}$), or falls below $\unit[11]{nm}$
while moving in the negative $x$-direction (for $v_x < \unit[0]{nm/ps}$), and
subsequently reenters the vapor phase within $\unit[30]{ps}$, the molecule is considered to have been
reflected from the liquid water slab. Additionally, the initial velocity of the single molecule and its angle of
incidence is recorded. The angle of incidence $\beta$ is defined as the angle
between the initial velocity vector and the surface normal of the water slab,
\emph{i.e.}, $\bm n = (1, 0, 0)$, see \Cref{fig:reflection} (a). This information allows for the calculation of the
reflection probability as
\begin{equation}
    p_\mathrm{ref} = \frac{N_\mathrm{ref}}{N_\mathrm{tot}}~,
\end{equation}
where $N_\mathrm{ref}$ is the number of reflected molecules within a given
velocity or angle range and $N_\mathrm{tot}$ is the total number of molecules in
that range. Only molecules with an initial velocity sufficiently high to leave
the vapor phase within the simulation time are considered when calculating $N_\mathrm{ref}$ and
$N_\mathrm{tot}$.

Each simulation run can be regarded as a Bernoulli trial, with possible outcomes
of 1 for a reflected molecule and 0 for an absorbed molecule. This framework
allows for the calculation of the statistical uncertainty of $p_\mathrm{ref}$:
For a given velocity or angle range (\emph{i.e.}, within a single bin), there are
$N_\mathrm{tot}$ independent runs, which leads to the standard error \cite{Durrett2019}
\begin{equation}
    \Delta p_\mathrm{ref} = \sqrt{\frac{p_\mathrm{ref} (1 - p_\mathrm{ref})}{N_\mathrm{tot}}}~.
    \label{eq:error}
\end{equation}

\subsection{Molecular reflection coefficient as a function of incidence angle and molecular velocity}
The results for the molecular reflection coefficient $p_\mathrm{ref}$ as a
function of the initial velocity of the impinging water molecule are shown in
\Cref{fig:reflection}(b), where the data from all three sets of simulations with
different initial velocity distributions are combined. 
The
reflection coefficient $p_\mathrm{ref}$ clearly increases with increasing initial
velocity of the impinging water molecule. Average velocities according to
$v=\sqrt{\kb T / m_\mathrm{w}}$ for the temperatures $T = \unit[300]{K}$, $\unit[1200]{K}$, and
$\unit[4800]{K}$ used in the simulations  are indicated by vertical dashed lines (from left to right). 
We observe that for a velocity corresponding to room temperature, $T$=300 K, $p_\mathrm{ref}$ indeed is negligibly small, as commonly assumed, while it rises significantly and reaches a value $p_\mathrm{ref}\approx 0.3$ for a velocity corresponding to a temperature of $T$=4800 K.
The reflection
coefficient  as a function of the angle of incidence $\beta$ of
the impinging water molecule is shown in Figs.~\ref{fig:reflection}(c)--(e) for
initial velocities drawn from Maxwell-Boltzmann distributions at
$\unit[300]{K}$, $\unit[1200]{K}$, and $\unit[4800]{K}$, respectively. We see that while for $T$=300K the reflection coefficient $p_\mathrm{ref}$ is rather independent of $\beta$, for $T$=1200 K and $T$=4800 K the reflection coefficient $p_\mathrm{ref}$ increases with increasing $\beta$ and this increase is more pronounced for the higher temperature.
This shows that, while the reflection coefficient
$p_\mathrm{ref}$ is small for water molecules impinging onto the liquid surface
with velocities typical for room temperature, it can become significant for
larger velocities and large angles of incidence.
Thus, the assumption of $p_\mathrm{ref} = 0$ used in \Cref{sec-sed-model}
is valid for a vapor phase that has a temperature close to room temperature, as relevant for droplet evaporation at ambient conditions.
\section{A multiphase diffuse-interface Cahn–Hilliard/Allen-Cahn model for evaporation and precipitation in droplets containing solutes}\label{sec-CH}
In this section we introduce an isothermal phase-field model for the coupled evaporation and crystallization of an aerosol droplet containing dissolved solutes: as the liquid evaporates, the dissolved solute, for example salt, becomes increasingly concentrated and, once it exceeds a saturation threshold, precipitates by forming a crystalline phase.
The model is described by a three-phase Cahn-Hilliard/Allen-Cahn system \cite{ch} with phase indicators $\varphi_i:[0,T]\times\Omega\to\R,$ where the index $i$ can represent a liquid phase $i=\ell$, a crystalline phase $i=c$, or a vapor phase $i=v$. The evolution of the phases is coupled with a diffusion equation for the solute concentration $s:[0,T]\times\Omega\to\R$. We assume that $s=1$ is the highest possible concentration (or volume fraction) of the pure crystal and $0\le s \le 1$.
A central modeling choice is to consider the dissolved or crystalline solute content as a conserved order parameter while allowing the liquid to become  supersaturated with solute. The saturation concentration is imposed softly, \emph{i.e.}, supersaturation is permitted but carries an energetic cost that increases smoothly beyond the saturation threshold.  
The focus of this section is to derive such a model based on a thermodynamic structure, to discuss suitable free energies, state-dependent mobilities, and reaction rates, and to study the properties of the model in numerical experiments. Our modeling approach is based on the works of Elliott \& Luckhaus \cite{EL} and Nestler \& Wheeler \cite{nestler2002phase}, Lamperti \& De Lorenzis \cite{Lamperti2026}. In this way, in this section we propose a three-dimensional, diffuse-interface model in terms of a three-phase Cahn-Hilliard/Allen-Cahn system to generalize the one-dimensional sharp-interface model discussed in \Cref{sec-sed-model}. Herein, the solute can be seen as a placeholder for any relevant substance, such as salt, proteins or viruses. 
In particular, with the numerical examples in \Cref{NumEx} we verify that the model is able to capture and generalize features of the one-dimensional model. 
The novelty of the present phase-field approach lies in combining evaporation, conserved solute transport, and precipitation or crystallization within a thermodynamically consistent three-phase gradient-flow framework and its application to evaporating solute droplets.
%
%
\subsection{Derivation of the thermodynamically consistent model}
\label{derivation}
%
\paragraph{Model derivation.}
We derive the thermodynamically consistent phase-field model by adapting the strategy proposed by Elliott and Luckhaus in \cite{EL} to our setting, where source terms in the order parameter equations and a coupling between the order parameter $\bs{\varphi}$ and solute concentration $s$ in the free energy density are taken into account.
\par
In a polyhedral, bounded domain $\Omega \subset \mathbb{R}^3$ and for times $t\in(0,T)$ we consider the state vector 
$\bs{q}=(\bs{\varphi},s)$, where $\bs{\varphi}=(\varphi_\ell,\varphi_c,\varphi_v)$ collects the phase-field functions that have to obey the additional constraint
\begin{align}
\label{eqn:constraint}
\varphi_\ell + \varphi_c + \varphi_v = 1\,.
\end{align}
Here $\varphi_i\approx 1$ means that the $i$th phase  is present and $\varphi_i\approx 0$ means that the $i$th phase is absent for $i\in\{\ell,c,v\}$.
The (isothermal) thermodynamics of these three phases $\bs\varphi$ and the solute concentration $s$ is modeled by a free energy functional $\mathcal{E}(\bs{q})=\int_\Omega \Psi\,\mathrm{d}x$ with free energy density $\Psi=\Psi(\bs{\varphi},\nabla\bs{\varphi},s)$ of Ginzburg-Landau type.  The above  constraint \Cref{eqn:constraint} for the phase indicators is taken into account with the aid of the Lagrangian functional $\mathcal{L}(\bs{q},\kappa)=\int_\Omega L\,\mathrm{d}x$ with density
\begin{equation}
\label{eq:Lagrangian}
L(\bs{\varphi},\nabla\bs{\varphi},s,\kappa)
:=\Psi(\bs{\varphi},\nabla\bs{\varphi},s)
+\kappa(\varphi_\ell+\varphi_c+\varphi_v-1)\,.    
\end{equation}
The driving forces of the processes in the droplet and the vapor are thus given by the chemical potentials $\bs{\mu}=(\mu_\ell,\mu_c,\mu_v,\mu_s)$ determined as the partial derivatives of the functional corresponding to the Lagrangian, \emph{i.e.},
\begin{align}
\label{eq:chempots}
\mu_\ell = \frac{\delta L}{\delta \varphi_\ell}
=\frac{\delta \Psi}{\delta \varphi_\ell}+\kappa\,, 
\qquad \mu_c = \frac{\delta L}{\delta\varphi_c}
=\frac{\delta \Psi}{\delta\varphi_c} + \kappa\,, 
\qquad \mu_v = \frac{\delta L}{\delta\varphi_v}
=\frac{\delta \Psi}{\delta\varphi_v}+\kappa\,,
\qquad \mu_s = \frac{\delta L}{\delta s}=\frac{\delta \Psi}{\delta s}\,.
\end{align}
\par
The mass balance equation for the order parameters 
$\bs{\varphi}=(\varphi_\ell,\varphi_c,\varphi_v)$ reads
\[
\frac{\mathrm{d}}{\mathrm{d}t} \int_\omega \varphi_i \, \mathrm{d}x
=
-\int_{\partial \omega} J_i \cdot \nu\,\mathrm{d}a(x) + \int_\omega M_i\, \mathrm{d}x\,,
 \qquad i \in \{\ell, c, v\}, \]
for any control volume $\omega\subset\Omega$ with  outer normal vector $\nu$, where we denote by $J_i$ the mass fluxes and by $M_i$ the source terms. 
Hence, it follows
\begin{equation}\label{massbalfhi}
    \partial_t \varphi_i = -\operatorname{div} J_i + M_i  \qquad i \in \{\ell, c, v\}.
\end{equation}
Since we are interested in the case where the solute concentration $s$ is conserved throughout the process, we postulate its mass balance equation to be of type
\begin{equation}\label{massbals}
\partial_t s = -\operatorname{div} J_s\,,
\end{equation}
for a mass flux $J_s$.

\noindent 
Let us consider the time derivative of the Lagrangian functional 
\begin{equation}
\label{enmu}
\begin{split}
\frac{\mathrm{d}}{\mathrm{d}t}\mathcal{L}(\bs{q},\kappa)
&=
\int_\Omega
\Big(
\mu_\ell \partial_t \varphi_\ell
+ \mu_c \partial_t \varphi_c+ \mu_v \partial_t \varphi_v
+ \mu_s\partial_t s
+(\varphi_\ell+\varphi_c+\varphi_v-1)\partial_t\kappa
\Big)\, \mathrm{d}x\\
&=
\int_\Omega
\Big(
\mu_\ell \partial_t \varphi_\ell
+ \mu_c \partial_t \varphi_c+ \mu_v \partial_t \varphi_v
+ \mu_s\partial_t s
\Big)\, \mathrm{d}x\,,
\end{split}
\end{equation}
where we used that the constraint is satisfied. 
By substituting \Cref{massbalfhi,massbals} 
 and imposing no-flux boundary conditions, we get
\begin{align*}
\int_\Omega \mu_i \partial_t\varphi_i dx &= \int_\Omega \left( \nabla \mu_i \cdot J_i + M_i \mu_i \right) \,\mathrm{d}x\,, \qquad i \in \{\ell, c, v\}\,,\\
\int_\Omega \mu_s \partial_t s \, \mathrm{d}x 
&= \int_\Omega \left( \nabla \mu_s\cdot J_s \right) \,\mathrm{d}x\,.
\end{align*}
Inserting these relations into \Cref{enmu}, we obtain
\begin{equation}
\label{enend}
\frac{\mathrm{d}}{\mathrm{d}t}\mathcal{L}(\bs{q},\kappa)
= \int_\Omega \sum_{i\in \{\ell,c,v,s\}} \nabla \mu_i \cdot J_i \, \mathrm{d}x
+\int_\Omega \sum_{i\in \{\ell,c,v\}} M_i \mu_i \, \mathrm{d}x\,.
\end{equation}

\noindent We now set $J_c=0$ (no diffusion of crystalline phase) and otherwise
\begin{align*}
J_i &:= - m_i(\bs{q}) \nabla \mu_i, \qquad i \in \{\ell, v,s\}\,,
\end{align*}
with state-dependent mobilities $m_i(\bs q) \ge 0$ for $i \in \{\ell, v, s\}$.  
Therefore, we infer

\begin{align*}
    \sum_{i\in \{\ell,v,s\}} \nabla \mu_i \cdot J_i &= - \sum_{i\in \{\ell,v,s\}} m_i(\bs{q}) |\nabla \mu_i|^2 \le 0\,.
\end{align*}

\noindent For the source terms in \Cref{massbalfhi}, we choose
\begin{align*}
M_\ell &:= h_\mathrm{cryst}(\bs{q})(\mu_c-\mu_\ell) + h_\mathrm{evap}(\bs{q})(\mu_v-\mu_\ell)\,,\\
M_c &:= h_\mathrm{cryst}(\bs{q})(\mu_\ell- \mu_c)\,,\\
M_v &:= h_\mathrm{evap}(\bs{q})(\mu_\ell-\mu_v)\,,
\end{align*}
with state-dependent reaction rates $h_\mathrm{cryst}(\bs{q}),h_\mathrm{evap}(\bs q)\geq 0$. Then, \Cref{enend} results in 
\begin{align}
\label{eqn:EDE}
\frac{\mathrm{d}}{\mathrm{d}t}\mathcal{L}\bigl(\bs{q}(t)\bigr)=-\int_\Omega m_\ell |\nabla \mu_\ell|^2 + m_v |\nabla \mu_v|^2 + m_s |\nabla \mu_s|^2 + h_\mathrm{evap} |\mu_\ell-\mu_v|^2 + h_\mathrm{cryst} |\mu_c-\mu_\ell|^2\,\mathrm{d}x\le 0\,,
\end{align}
which proves that the Lagrangian functional related to the free energy of the system decreases in time, and hence that the  model is thermodynamically consistent. 

\paragraph{The resulting PDE-system.}
In summary, with the above choices for the source terms and fluxes the evolution laws \Cref{massbalfhi,massbals} describing the diffusion, evaporation and crystallization processes in the solution droplet and the vapor phase thus result in the following coupled PDE-system in $(0,T)\times\Omega$
\begin{subequations}
\label{eqn:evo}
\begin{align}
\label{eqn:evo-phil}
\partial_t \varphi_\ell - \operatorname{div} (m_\ell(\bs{q}) \nabla \mu_\ell) &= h_\mathrm{cryst}(\bs{q})(\mu_c-\mu_\ell) + h_\mathrm{evap}(\bs{q})(\mu_v-\mu_\ell) \,,\\
\partial_t \varphi_c &= h_\mathrm{cryst}(\bs{q})(\mu_\ell-\mu_c)\,, \\
\partial_t \varphi_v - \operatorname{div}(m_v(\bs{q})\nabla\mu_v)&=h_\mathrm{evap}(\bs{q})(\mu_\ell-\mu_v)\,,\\
\label{eqn:evo-s}
\partial_t s - \operatorname{div} (m_s(\bs{q}) \nabla \mu_s) &= 0\,,\\
\label{eqn:evo-const}
\varphi_\ell + \varphi_c + \varphi_v &= 1\,,
\end{align}
\end{subequations}
complemented with no-flux boundary conditions 
and with an initial condition $\bs{q}(t=0)=\bs{q}^0$. The mobilities $m_\ell,m_v,m_s\ge 0$ and reaction rates $h_\mathrm{evap},h_\mathrm{cryst}\ge 0$ for evaporation and crystallization/precipitation are state-dependent functions. Their choice as well as the choice of the free energy density $\Psi$ shall be specified in more detail below in \Cref{DiscussFreeEnDiss}.
\paragraph{Gradient structure of \Cref{eqn:evo}.} Following \emph{e.g.}\ \cite{Miel13TMER,ZHPVJT23PMMR}, it can be observed that system \Cref{eqn:evo} has a gradient structure. For this, we introduce the dual dissipation potential 
$\mathcal{D}^*(\bs{q};\bs{\mu}):=\mathcal{D}_\mathrm{D}^*(\bs{q};\bs{\mu})+\mathcal{D}_\mathrm{R}^*(\bs{q};\bs{\mu})$ with $\mathcal{D}_\mathrm{D}^*(\bs{q};\bs{\mu}):=\int_\Omega D^*_{\mathrm{D}}(\bs{q};\nabla\bs{\mu})\,\mathrm{d}x$ and 
$\mathcal{D}_\mathrm{R}^*(\bs{q};\bs{\mu}):=\int_\Omega D^*_{\mathrm{R}}(\bs{q};\bs{\mu})\,\mathrm{d}x$ the dual dissipation potentials for the diffusion and the reaction processes with densities 
$D^*(\bs{q};\bs{\mu},\nabla\bs{\mu}):=
D^*_{\mathrm{D}}(\bs{q};\nabla\bs{\mu})+
D^*_{\mathrm{R}}(\bs{q};\bs{\mu}),$ 
\begin{equation*}
\begin{split}
D^*_{\mathrm{D}}(\bs{q};\nabla\bs{\mu})&:=\tfrac{1}{2}\nabla\bs{\mu}:\mathbb{M}_\mathrm{D}(\bs{q})\nabla\bs{\mu}\,,\quad\text{where }\;  
\mathbb{M}_{\mathrm{D}}(\bs{q}):=
\mathrm{diag}\big(m_\ell(\bs{q}),0,m_v(\bs{q}),m_s(\bs{q})\big)\,,
\quad\text{and}\quad\\
D^*_{\mathrm{R}}(\bs{q};\bs{\mu})&:=\tfrac{1}{2}\Big(
h_\mathrm{evap}(\bs{q})|\mu_\ell-\mu_v|^2
+h_\mathrm{cryst}(\bs{q})|\mu_c-\mu_\ell|^2
\Big)\,
\end{split}
\end{equation*}
for $\bs{\mu}=(\mu_\ell,\mu_c,\mu_v,\mu_s)$. 
It is easy to see that the potentials are quadratic and positively semidefinite with respect to the variable $\bs{\mu}$, so that their functional derivatives result in symmetric and positively semidefinite operators.  
Moreover, one finds that system 
\Cref{eqn:evo} is given by the evolution law
\begin{equation}
\label{gradsyst}
\partial_t\bs{q}
=\mathrm{D}_{\bs{\mu}}\mathcal{D}^*(\bs{q};-\mathrm{D}_{\bs{q}}\mathcal{L}(\bs{q},\kappa))\,. 
\end{equation}
Testing the gradient system \Cref{gradsyst} by $\bs{\mu}=\mathrm{D}_{\bs{q}}\mathcal{L}(\bs{q},\kappa))$ results in 
\begin{equation}
\label{EDE1}
\begin{split}
\frac{\mathrm{d}}{\mathrm{d}t}\mathcal{L}(\bs{q},\kappa)=\langle\mathrm{D}_{\bs{q}}\mathcal{L}(\bs{q},\kappa),\partial_t\bs{q}\rangle
&=
\langle\mathrm{D}_{\bs{q}}\mathcal{L}(\bs{q},\kappa),\mathrm{D}_{\bs{\mu}}\mathcal{D}^*(\bs{q};-\mathrm{D}_{\bs{q}}\mathcal{L}(\bs{q},\kappa))\rangle\\
&=-\int_{\Omega}\Big(
\nabla\bs{\mu}:\mathbb{M}_\mathrm{D}(\bs{q})\nabla\bs{\mu}
+
h_\mathrm{evap}(\bs{q})|\mu_\ell-\mu_v|^2
+h_\mathrm{cryst}(\bs{q})|\mu_c-\mu_\ell|^2
\Big)\,\mathrm{d}x\\
&=-2\Big(\mathcal{D}^*_\mathrm{R}(\bs{q};\bs{\mu})+\mathcal{D}^*_\mathrm{D}(\bs{q};\bs{\mu})\Big)\leq0\,,
\end{split}
\end{equation}
which is the energy-dissipation estimate \Cref{eqn:EDE} recovered from the gradient structure. In order to address the different dissipative processes separately we also introduce
\begin{align}
\mathcal{D}^*_\mathrm{cryst}:=\int_\Omega\tfrac{1}{2}h_\mathrm{cryst}(\mu_c-\mu_\ell)^2\,\mathrm{d}x, \quad \mathcal{D}^*_\mathrm{evap}:=\int_\Omega \tfrac{1}{2}h_\mathrm{evap}(\mu_v-\mu_\ell)^2\,\mathrm{d}x, \quad \mathcal{D}^*_{m_i}:= \int_\Omega \tfrac{1}{2}m_i|\nabla\mu_i|^2\,\mathrm{d}x\,,
\end{align}
for $i\in\{s,\ell,v\}$ such that $\mathcal{D}^*_\mathrm{D}=\mathcal{D}^*_{m_\ell}+\mathcal{D}^*_{m_v}+\mathcal{D}^*_{m_s}$ and $\mathcal{D}^*_\mathrm{R}=\mathcal{D}^*_\mathrm{cryst}+\mathcal{D}^*_\mathrm{evap}$. For the free energy this gives the  balance
\begin{align}
\mathcal{E}\bigl(\bs{q}(t)\bigr)-\mathcal{E}\bigl(\bs{q}(0)\bigr)=-\int_0^t 2\mathcal{D}^*\bigl(\bs{q}(\tau);\bs{\mu}(\tau)\bigr)\,\mathrm{d}\tau\,,
\end{align}
which we use later to qualify the relevance of different dissipative processes in a time-integrated manner.

%
\subsection{Weak formulation and saddle point structure}
\label{WeakF}

\paragraph{Weak formulation.}
We rewrite system \Cref{eqn:evo} in terms of a weak formulation, where we seek $(\bs{q},\bs{\mu},\kappa)$ as unknown functions with corresponding test functions $(\bs{w},\bs{\xi},w_\kappa)$ with components $\bs{w}=(w_\ell,w_c,w_v,w_s)$ and $\bs{\xi}=(\xi_\ell,\xi_c,\xi_v,\xi_s)$. For a.a.\ $t\in(0,T)$ we thus seek $(\bs{q}(t),\bs{\mu}(t),\kappa(t))$ such that
\begin{subequations}
\label{eqn:weakform}
\begin{align}
\nonumber
\int_\Omega \Big[m_\ell(\bs{q}) \nabla\mu_\ell\cdot\nabla \xi_\ell + \xi_\ell \partial_t \varphi_\ell + m_s(\bs{q}) \nabla\mu_s \cdot \nabla\xi_s + \xi_s \partial_t s  + m_v(\bs{q})\nabla\mu_v\cdot\nabla\xi_v + \xi_v\partial_t\varphi_v\Big.\\
\label{eqn:weakform-1}
\Big.+h_\mathrm{evap}(\bs{q}) (\mu_\ell-\mu_v)(\xi_\ell-\xi_v)  + h_\mathrm{cryst}(\bs{q})(\mu_c-\mu_\ell)(\xi_c-\xi_\ell) + \xi_c \partial_t \varphi_c\Big] \,\mathrm{d}x &= 0\,,\\
\label{eqn:weakform-2}
\int_\Omega \mu_\ell w_\ell + \mu_s w_s + \mu_c w_c+\mu_v w_v\,\mathrm{d}x - \langle \mathrm{D}\mathcal{L}(\bs{q},\kappa),(\bs{w},w_\kappa)\rangle &=0\,,
\end{align}
\end{subequations}
for all $(\bs{w},w_\kappa,\bs{\xi})$, 
where we abbreviated the Fr\'echet derivative of the Lagrange functional in the direction $(\bs{w},w_\kappa)$ by 
\begin{align}
\langle \mathrm{D}\mathcal{L}(\bs{q},\kappa), (\bs{w},w_\kappa) \rangle=\int_\Omega \partial_s \Psi\,w_s + \!\!\!\!\!\sum_{i\in\{\ell,c,v\}}\!\!\!\!\!\big(\partial_{\varphi_i} \Psi \cdot w_i + (\partial_{\nabla\varphi_i}\Psi) \cdot \nabla w_i + \kappa w_i\big) + w_\kappa(\varphi_c+\varphi_\ell+\varphi_v-1)\,\mathrm{d}x\,.
\end{align}
In this way, \Cref{eqn:weakform-2} ensures relations \Cref{eq:chempots} as well as the constraint \Cref{eqn:evo-const}, whereas \Cref{eqn:weakform-1} comprises the weak form of the evolution laws \Cref{eqn:evo}. 
Observe that, by testing \Cref{eqn:weakform-2} with $\bs{w}=\partial_t \bs{q}$, $w_\kappa=\partial_t\kappa$ and \Cref{eqn:weakform-1} with $\bs{\xi}=\bs{\mu}$ we obtain the energy-dissipation estimate \Cref{eqn:EDE}.
\paragraph{Saddle point structure.}
Exploiting the gradient structure \Cref{gradsyst}, we can introduce the bilinear forms 
\begin{subequations}
\begin{align}
\nonumber
a(\bs{\mu},\bs{\xi})&:=\langle\mathrm{D}_{\bs{\mu}}
\mathcal{D}^*(\bs{q};\bs{\mu}),\bs{\xi}\rangle
\\
&\phantom{:}=\int_\Omega \Big[ \sum_{i\in\{\ell,v,s\}}\Big[m_i \nabla\mu_i\cdot\nabla \xi_i\Big] + h_\mathrm{evap} (\mu_\ell-\mu_v)(\xi_\ell-\xi_v)  + h_\mathrm{cryst}(\mu_c-\mu_\ell)(\xi_c-\xi_\ell) \Big] \,\mathrm{d}x\,,\\
b(\bs{w},\bs{\mu})&:=\int_\Omega \Big(\mu_\ell w_\ell + \mu_s w_s + \mu_c w_c+\mu_v w_v\Big)\,\mathrm{d}x\,,
\end{align}
\end{subequations}
and rewrite the weak formulation \Cref{eqn:weakform} in the following saddle-point structure
\begin{subequations}
\label{saddle}
\begin{align}
\label{saddle-1}
a(\bs{\mu},\bs{\xi}) + b(\partial_t \bs{q},\bs{\xi}) &= 0\,,\\
\label{saddle-2}
b(\bs{w},\bs{\mu}) &= \langle \mathrm{D}\mathcal{L}(\bs{q},\kappa),(\bs{w},w_\kappa)\rangle\,,
\end{align}
\end{subequations}
for all $(\bs{w},w_\kappa,\bs{\xi})$. In this way \Cref{saddle-1} coincides with  \Cref{eqn:weakform-1} and \Cref{saddle-2} 
with \Cref{eqn:weakform-2}. 
Repeating the test with $\bs{w}=\partial_t\bs{q}$, $w_\kappa=\partial_t\kappa$, $\bs{\xi}=\bs{\mu}$ in \Cref{saddle} we get 
\begin{align}
\frac{\rm d}{\mathrm{d}t}\mathcal{L}(\bs{q}(t))=\langle \mathrm{D}\mathcal{L}(\bs{q},\kappa),(\partial_t\bs{q},\partial_t\kappa)\rangle=b(\partial_t\bs{q},\bs{\mu})=-a(\bs{\mu},\bs{\mu}) = -2\mathcal{D}^*(\bs{q};\bs{\mu}) \le 0\,,
\end{align}
which is again \Cref{eqn:EDE}, recovered from the gradient structure of the coupled system, as in \Cref{EDE1}. 

%
\subsection{Choice of free energy and dissipation potentials}
\label{DiscussFreeEnDiss}

\begin{figure}
\includegraphics[height=2.4in]{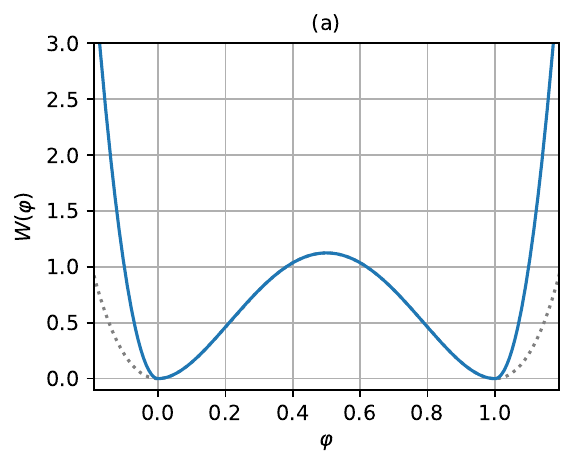}\qquad\qquad%
\includegraphics[height=2.4in]{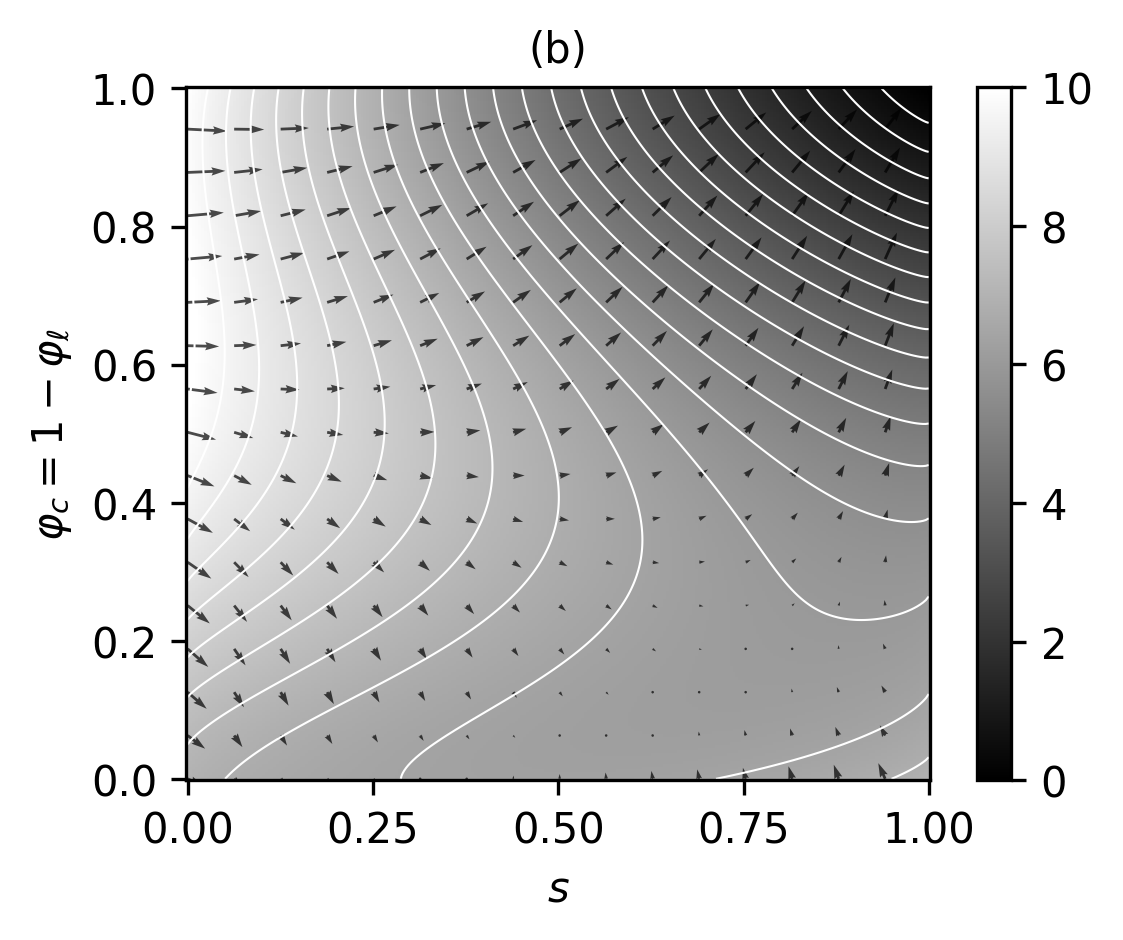}
\caption{(a) Potential $W(\varphi)$ from \Cref{choice-W} for $\Lambda=100$ (solid, blue) compared to standard quartic $18\varphi^2(1-\varphi)^2$ (dotted, gray)  and (b) energy landscape $(\Psi-\min\Psi)$ with $\lambda=10$, $\beta=-10$, $\Lambda=100$, $\gamma_\ell=\gamma_c=1/8$, $\gamma_v=2$, $\varepsilon=0.2$, $s_\mathrm{sat}=0.3$ for a homogeneous solution (no gradients) without vapor $\varphi_v=0$, \emph{i.e.}, $\varphi_c=1-\varphi_\ell$ with isolines and negative gradient vector field.}
\label{fig:Wfunction}
\end{figure}

In the following we specify in more detail a choice for the free energy density and the state-dependent mobilities and reaction rates suited to capture certain effects as evaporation and crystallization progress in the droplet. 
\paragraph{Free energy.}
As discussed in \Cref{derivation}, we consider the free energy and the Lagrange functional 
\begin{align}
\mathcal{E}(\bs{q})
:= 
\int_\Omega \Psi(\bs{\varphi}, \nabla\bs{\varphi}, s)\,\mathrm{d}x\,,
\qquad\qquad 
\mathcal{L}(\bs{q},\kappa):=\mathcal{E}(\bs{q}) + \int_\Omega \kappa (\varphi_\ell + \varphi_c + \varphi_v - 1)\,\mathrm{d}x\,.
\end{align}
Now we set $\Psi$ as follows 
\begin{equation}
\label{choice-Psi}
\begin{split}
&\Psi(\bs{\varphi}, \nabla\bs{\varphi}, s):= 
\sum_{i\in \lbrace \ell, c, v\rbrace} \gamma_i\left[ \frac{\varepsilon}{2} |\nabla \varphi_i|^2
+  \frac{1}{\varepsilon}W(\varphi_i)\right]
+ \Pi(\bs{\varphi}, s)\,,\\
&\Pi(\bs{\varphi},s):=s\ln(s) + (1-s)\ln(1-s) + \varphi_c \beta (s-s_\mathrm{sat}) + \lambda \varphi_v s\,,
\end{split}
\end{equation}
where $\varepsilon$ denotes the thickness of the interface, $\gamma_i$ represent the surface tension coefficients, $W$ is the phase-field potential and the potential $\Pi$ takes into account the coupling between $\bs{\varphi}$ and $s$. 
More precisely, in $\Pi$ we have the free energy that drives the solute diffusion and keeps $0<s<1$ as well as an extra term that for $\beta<0$ favors crystallization beyond a saturation threshold, \emph{i.e.}, for $s>s_\mathrm{sat}$. 
The classical mixture term with sufficiently large $\lambda>0$ prevents the solute from entering the vapor phase.
Note that the standard quartic double well does {not} strongly  enforce the condition $0\le \varphi_i\le 1$, for which other phase-field energies of logarithmic or double-obstacle type might be better suited (see \cite{ggps-ch,gp-ac} and references therein). Therefore, in order to better capture the constraint $0\le \varphi_i\le 1,$ the phase-field energy $W\in\mathrm{C}^1(\R)$ is chosen in the form
\begin{align}
\label{choice-W}
W(\varphi):=\begin{cases} \Lambda\varphi^2 & \varphi<0\,,\\
18\varphi^2(1-\varphi)^2 & 0\le \varphi \le 1\,,\\
\Lambda(\varphi-1)^2 & \varphi>1\,,
\end{cases}
\end{align}
for a sufficiently large $\Lambda\gg 1$ to additionally penalize values of $\varphi$ outside the interval $[0,1]$, see \Cref{fig:Wfunction}.
The terms multiplying $\gamma_i$ encode the surface tension between the $i$th and the $j$th phase via 
\begin{equation}
\label{surftens}
\gamma_{ij}=\tfrac12(\gamma_i+\gamma_j)>0
\quad\text{ for $i,j\in \{\ell,c,v\}$ and $i\neq j$.}
\end{equation}
 Note that this gives the three independent coefficients determined by the values $\gamma_{\ell v}, \gamma_{\ell c}$, $\gamma_{cv}$, where not all values might be feasible due to the restriction $\gamma_i\ge 0$ of this particular phase-field energy. 
\paragraph{Chemical potentials.}
With the above choice of free energy and Lagrangian functional we observe that the chemical potentials from \Cref{eq:chempots} take the specific form 
\begin{equation}\label{chempot}
\begin{aligned}
\mu_\ell
=\frac{\delta L}{\delta\varphi_\ell}=&
-\gamma_\ell \varepsilon \Delta \varphi_\ell
+ \frac{\gamma_\ell}{\varepsilon}  \frac{\partial W}{\partial\varphi_\ell}
+ \frac{\partial \Pi}{\partial\varphi_\ell}+\kappa\,, \\[3pt]
\mu_c
=\frac{\delta L}{\delta\varphi_c}=&
-\gamma_c \varepsilon \Delta \varphi_c
+ \frac{\gamma_c}{\varepsilon}  \frac{\partial W}{\partial\varphi_c}
+ \frac{\partial \Pi}{\partial \varphi_c}+\kappa\,,\\[3pt]
\mu_v= \frac{\delta L}{\delta\varphi_v}=&
-\gamma_v \varepsilon \Delta \varphi_v
+ \frac{\gamma_v}{\varepsilon}  \frac{\partial W}{\partial\varphi_v}
+\frac{\partial \Pi}{\partial \varphi_v}+\kappa\,,\\
\mu_s=\frac{\delta L}{\delta s}\;=&\,\frac{\partial\Pi}{\partial s}\,.
\end{aligned}
\end{equation}
\paragraph{Mobilities and reaction rates.}
For the liquid, vapor, and solute diffusive mobilities $m_\ell,$ $m_v,$ and $m_s$ we use 
\begin{align}
\label{choice-m}
\begin{split}
&m_\ell(\bs{q}):=\bigl(m_{\ell\ell}|\varphi_\ell| + m_{\ell v}|\varphi_v| + m_{\ell c}|\varphi_c|\bigr)\,,\\
&m_v(\bs{q}):=\bigl(m_{v\ell}|\varphi_\ell| + m_{vv}|\varphi_v| + m_{vc}|\varphi_c|\bigr)\,,\\
&m_{s}(\bs{q}):=\bigl(m_{s\ell}|\varphi_\ell| + m_{s v}|\varphi_v| + m_{s c}|\varphi_c|\bigr)s(1-s)\,,
\end{split}
\end{align}
for some given constant $m_{ij}>0$ that set the value of the liquid, vapor, and solute mobility in the pure liquid, vapor or crystalline phase for $i\in\{\ell,v,s\}$ and $j\in\{\ell,v,c\}$, respectively. Note that the form of $m_s$ makes sure that diffusion of the solute is confined to regions where $s\in(0,1)$. For the reaction rates we use 
\begin{align}
\label{choice-h}
h_\mathrm{evap}(\bs{q}):=h_{e}^0|\varphi_\ell\varphi_v|\,,\qquad
h_\mathrm{cryst}(\bs{q}):=h_{c}^0|\varphi_\ell|\,,
\end{align}
for some given constant $h_{\ell}^0,h_{c}^0>0$. The form of the evaporation rate makes sure that evaporation is restricted to the liquid-vapor interface whereas the crystallization rate restricts crystallization to the presence of a fluid phase. 
%

%
\subsection{Numerical examples}
\label{NumEx}

In the following we carry out numerical simulations using the three-phase model \Cref{eqn:evo}, also making use of the specific form of the potentials discussed in \Cref{DiscussFreeEnDiss}. Based on the weak formulation introduced in \Cref{WeakF} we provide a  discrete scheme in space and time in \Cref{Sec:discr-st}.  Subsequently, in \Cref{Sec:DropletEx} we present and discuss numerical results for evaporating droplets in different scenarios by varying certain parameter sets, such as the initial solute concentration and the parameters $\lambda$  and $\beta$ in the choice of the free energy \Cref{choice-Psi}. The Python code used in this section with the corresponding example parameters is published in \cite{Peschka2026EvaporatingDroplet}. 
\subsubsection{Discretization in space and time.}
\label{Sec:discr-st}
In order to discretize the weak formulation \Cref{eqn:weakform} in time we introduce $0=t^0<t^1<\ldots<t^N=T$ and, for each $k\in\{1,\ldots,N\}$ we set $\bs{q}^k:=\bs{q}(t^k)=(\varphi_\ell^k,\varphi_c^k,\varphi_v^k,s^k)$ and correspondingly $\bs{\mu}^k:=\bs{\mu}(t^k)=(\mu^k_\ell,\mu^k_c,\mu^k_v,\mu^k_s)$ and the multiplier $\kappa^k:=\kappa(t^k)$. For each $k\in\{1,\ldots,N\}$ we seek $(\bs{q}^k,\bs{\mu}^k,\kappa^k)$ such that
\begin{subequations}
\label{eqn:disc_weakform}
\begin{align}
\nonumber
\int_\Omega m_\ell^{k-1} \nabla\mu^k_\ell\cdot\nabla \xi_\ell + \xi_\ell\left(\tfrac{\varphi_\ell^k- \varphi_\ell^{k-1}}{\tau^k}\right)  +
m_s^{k-1} \nabla\mu^k_s \cdot \nabla\xi_s + \xi_s \left(\tfrac{s^k- s^{k-1}}{\tau^k}\right)+
m_v^{k-1} \nabla\mu^k_v \cdot \nabla\xi_v + \xi_v \left(\tfrac{\varphi_v^k- \varphi_v^{k-1}}{\tau^k}\right)&
\\
\label{discr-1}
+\,\,\xi_c\left(\tfrac{\varphi_c^k- \varphi_c^{k-1}}{\tau^k}\right) + h^{k-1}_\mathrm{evap}(\mu^k_v-\mu^k_\ell)(\xi_v-\xi_\ell) + h^{k-1}_\mathrm{cryst}(\mu_c^k-\mu_\ell^k)(\xi_c-\xi_\ell)  \,\mathrm{d}x &= 0\,,\\
\label{discr-2}
\int_\Omega \mu^k_\ell w_\ell + \mu^k_s w_s + \mu^k_c  w_c + \mu^k_v w_v \,\mathrm{d}x - \langle \mathrm{D}\mathcal{L}(\bs{q}^k,\kappa^k), (\bs{w},w_\kappa)\rangle&=0\,,
\end{align}
\end{subequations}
for all $(\bs{w},w_\kappa,\bs{\xi})$. Above we abbreviated $m_i^k:=m_i(\bs{q}^k)$ and $h_\alpha^k:=h_\alpha(\bs{q}^k)$ for $i\in\{\ell,s,v\}$ and $\alpha\in\{\mathrm{evap},\mathrm{cryst}\}$ and $\tau^k:=t^k-t^{k-1}$. This nonlinear saddle point problem we discretize in space via $P^1$ finite elements for  $\bs{q}^k$, $\bs{\mu}^k$ and $\kappa^k$ and solve it via Newton's method. 
Due to the explicit handling of the mobilities, the nonsmooth state-dependence via $|\varphi^{k-1}_i|$ terms is unproblematic for the Newton solver. 
For discretization and solution we use the finite element framework FEniCS \cite{logg2012automated}. We employ an adaptive time step control based on the number of Newton steps per iteration to reach a specified tolerance of the residual. We implement a spherically symmetric setup with radial coordinate $r=\sqrt{x_1^2+x_2^2+x_3^2}=|\bs{x}|$ and $\bs{x}=(x_1,x_2,x_3)\in\Omega$ by replacing in the weak formulation \Cref{eqn:disc_weakform} all integrals as follows
$
\int_\Omega \ldots\,\,\mathrm{d}x \quad\to\quad \int_0^L \ldots \,\,r^2\,\mathrm{d}r
$
to place a spherical droplet of radius $R_0<L$.
%
\subsubsection{Examples for evaporating droplets}
\label{Sec:DropletEx}

In the following, we present and discuss parameter sets shown in  \Cref{tab:params} for a droplet of initial size $R_0=3$ and a domain of radius $L=4$. Throughout the examples, we vary some selected parameters as specified below in \Cref{tab:varparams}. We use the initial data 
\begin{equation}
\label{initial-data}
\varphi_c^0(r)=0\,,\qquad 
\varphi_v^0(r)=\tfrac12 \left[1+\tanh\left(\tfrac{3}{\varepsilon}(r-R_0)\right)\right]\,, \qquad 
\varphi_\ell^0(r)=1-\varphi_v^0(r)\,,\qquad 
s^0(r)= \bar{s}^0\exp(-\lambda\varphi^0_v(r))\,,
\end{equation}
that encode an initial liquid droplet for $r<R_0$ with an adjacent vapor phase for $r>R_0$ but no initial crystalline phase. The initial solute concentration in the liquid is $\bar{s}^0$.  We solve problem \Cref{eqn:evo} for $0<t<T$. The main idea of the following spherically symmetric 3D examples is to drive evaporation via the vapor surface tension, mainly  through $\gamma_v=2$,  with smaller values $\gamma_\ell=\gamma_c=1/8$. The main modifications in the examples are the solute concentration $\bar{s}^0$ in \Cref{initial-data}  and the parameters $\lambda$ and $ \beta$ in the free energy density 
\Cref{choice-Psi}. The role of $\lambda$ is to energetically penalize the solute from entering the vapor phase, whereas $\beta$ introduces a tilt to the phase-field energy that, for sufficiently negative values, favors the creation of a crystalline phase. 
\begin{table}[hb!]
\caption{General parameters of energy and dissipation 
\label{tab:params}}
\begin{ruledtabular}
\begin{tabular}{lcccccccccccc}
\textbf{parameter}
& $\gamma_\ell$ & $\gamma_v$ & $\gamma_c$ & $\varepsilon$ & $s_\mathrm{sat}$
& $\Lambda$
& $(m_{\ell\ell},m_{\ell v},m_{\ell c})$
& $(m_{v\ell},m_{vv},m_{vc})$
& $(m_{s\ell},m_{sv},m_{sc})$
& $h_e^0$ & $h_c^0$ \\
\textbf{value}
& $1/8$ & $2$ & $1/8$ & $0.2$ & $0.3$ 
& $10^{2}$
& $(1,1,10^{-2})$ & $(10^{-2},1,10^{-2})$ & $(10^{-2},1,10^{-2})$ & 1  & 1 \\
\end{tabular}
\end{ruledtabular}
\end{table}
\begin{table}[hb!]
\caption{Varying parameters throughout Examples (a)--(d) \label{tab:varparams}}
\begin{ruledtabular}
\begin{tabular}{lccc}
\textbf{parameter}
& $\lambda$ & $\beta$ & $\bar s^0$  \\
\textbf{value Ex.\ (a)}
& $1$ & $-1$ & $10^{-2}$  \\
\textbf{value Ex.\ (b)}
& $10$ & $-1$ & $10^{-1}$  \\
\textbf{value Ex.\ (c)}
& $10$ & $-10$ & $10^{-1}$  \\
\textbf{value Ex.\ (d)}
& $10$ & $-10$ & $10^{-2}$  \\
\end{tabular}
\end{ruledtabular}
\end{table}

Note that the parameters in use are not calibrated to any experimental data. They are chosen in such a way as to illustrate their qualitative influence on the droplet evolution.

\paragraph*{Example (a): Droplet completely evaporates.}

\begin{figure}[H]
\centering
\includegraphics[width=0.75\textwidth]{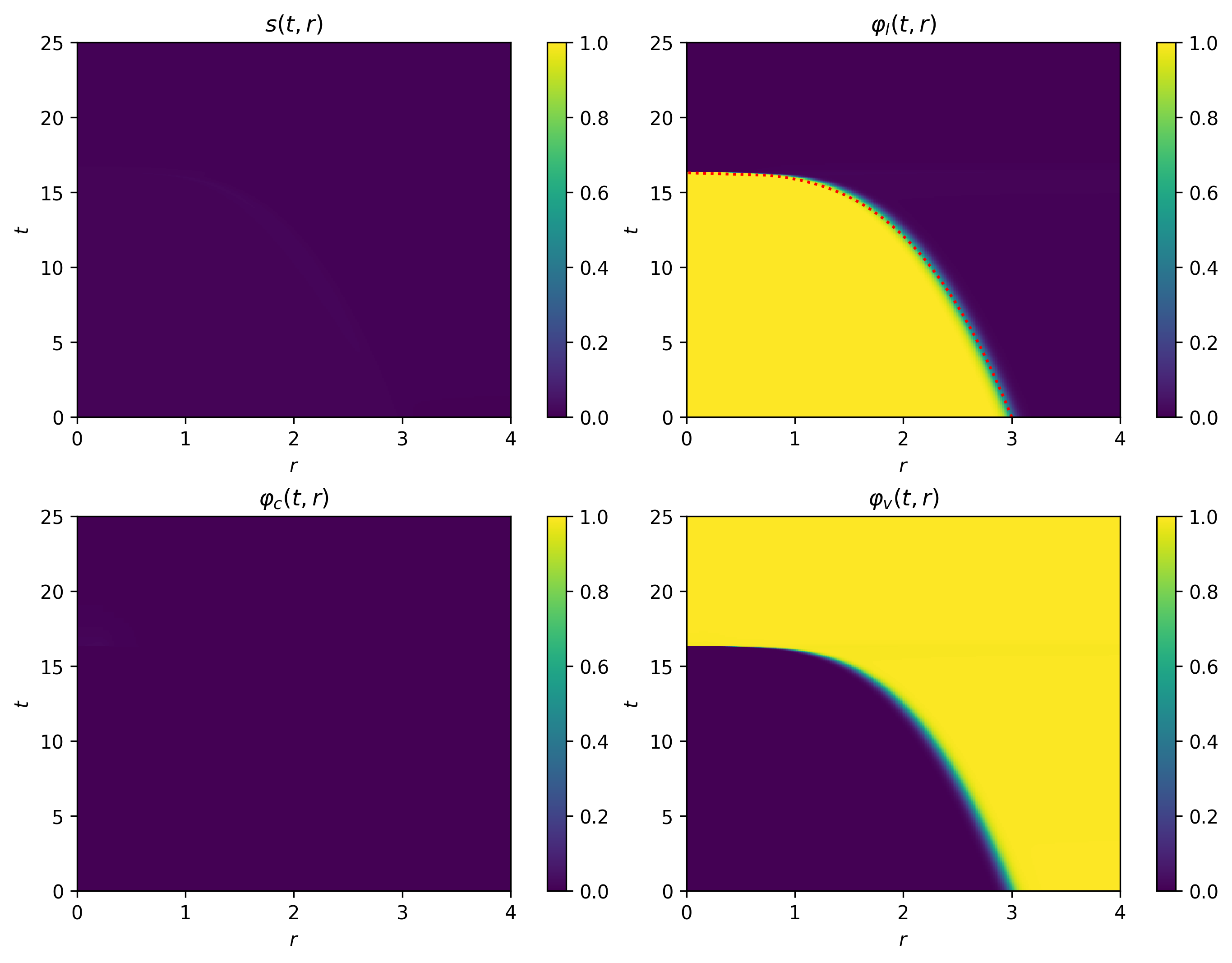}
\caption{Solute concentration $s(t,r)$ and phase fields $\varphi_i(t,r)$ for $i=\{\ell,c,v\}$ as a function of time $t$ and radius $r$ for different parameters for Example (a) with $\lambda=1$, $\beta=-1$, $\bar{s}^0=10^{-2}$, where the droplet completely evaporates. The dotted red curve indicates the function $R(t)= (R_0^{1/\alpha}-Ct)^\alpha$ for $\alpha=0.3$.}
\label{fig:examples_cryst_a}
\end{figure}

\begin{figure}[ht]
\centering
\includegraphics[width=0.42\textwidth]{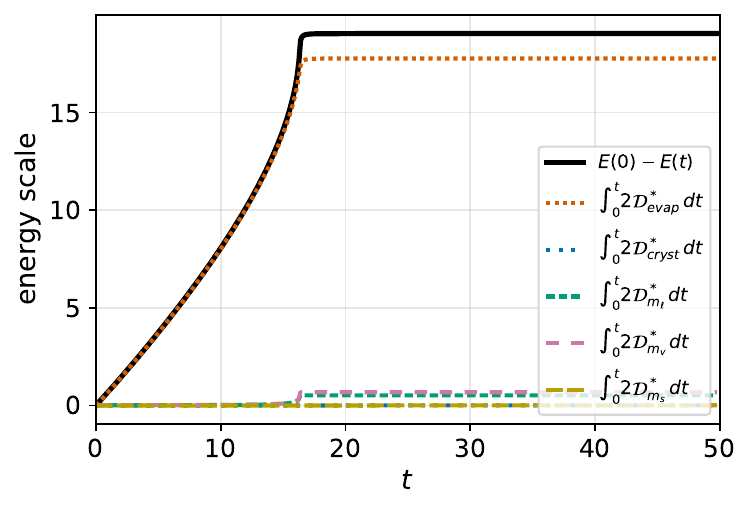}\qquad
\includegraphics[width=0.42\textwidth]{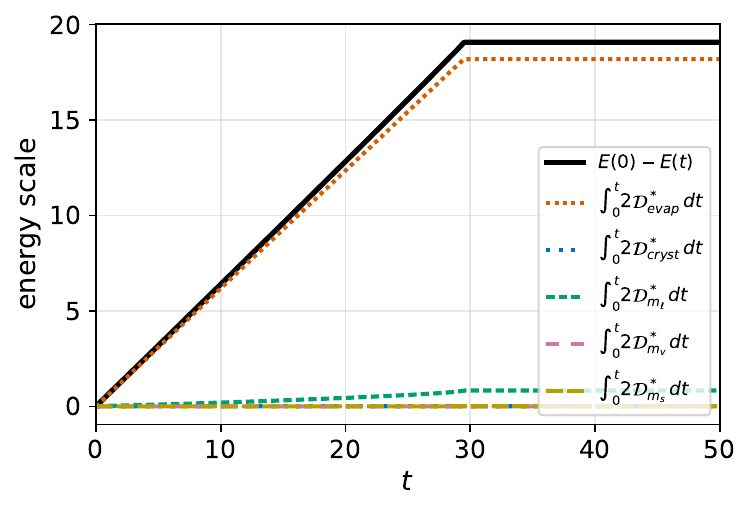}
\caption{Energy $\mathcal{E}$ and different time-integrated contributions to the dissipation $\mathcal{D}^*$ for Example (a) shown in \Cref{fig:examples_cryst_a} and \Cref{fig:examples_cryst_a1} (\textbf{left}) with the mobilities from \Cref{tab:params} and (\textbf{right}) with the reduced liquid mobilities $m_{\ell i}=10^{-3}$ for $i\in\{\ell,v,c\}$.}
\label{fig:energyX_a} 
\end{figure}

For the first example we select a low solute concentration $\bar{s}^0=10^{-2}$ and moderately large values $\lambda=1$, $\beta=-1$, so that the solute can eventually be dispersed in the vapor phase and the crystalline phase is not strongly favored energetically. The corresponding solution of the phase field model is shown in \Cref{fig:examples_cryst_a} 
over the time interval $[0,25]$. Starting with a droplet of initial radius $3$, the droplet size shrinks over time as can be seen in the
plot for $\varphi_\ell$. 
As the radius has shrunk to the size of $R\approx 1.5$ at time $t=16$ the evaporation process rapidly accelerates and leads to the  extinction of the droplet at time $t\approx 16$. 
In accordance with the low solute concentration, no crystalline phase is formed, cf.\ the plot for $\varphi_c$ in  \Cref{fig:examples_cryst_a}. 
Accordingly, the vapor phase, depicted in 
the plot of $\varphi_v$ in  \Cref{fig:examples_cryst_a}, 
is the complement of the liquid phase. The evolution of the free energy and time-integrated dissipation during this process is depicted in the left panel of  \Cref{fig:energyX_a}. One can see that the free energy rapidly decreases due to evaporation, which is the main dissipative process, until the droplet is extinguished at $t\approx 16$ close to the steady state, after which the energy remains (approximately) constant. 

\begin{figure}[ht]
\centering
\includegraphics[width=0.75\textwidth]{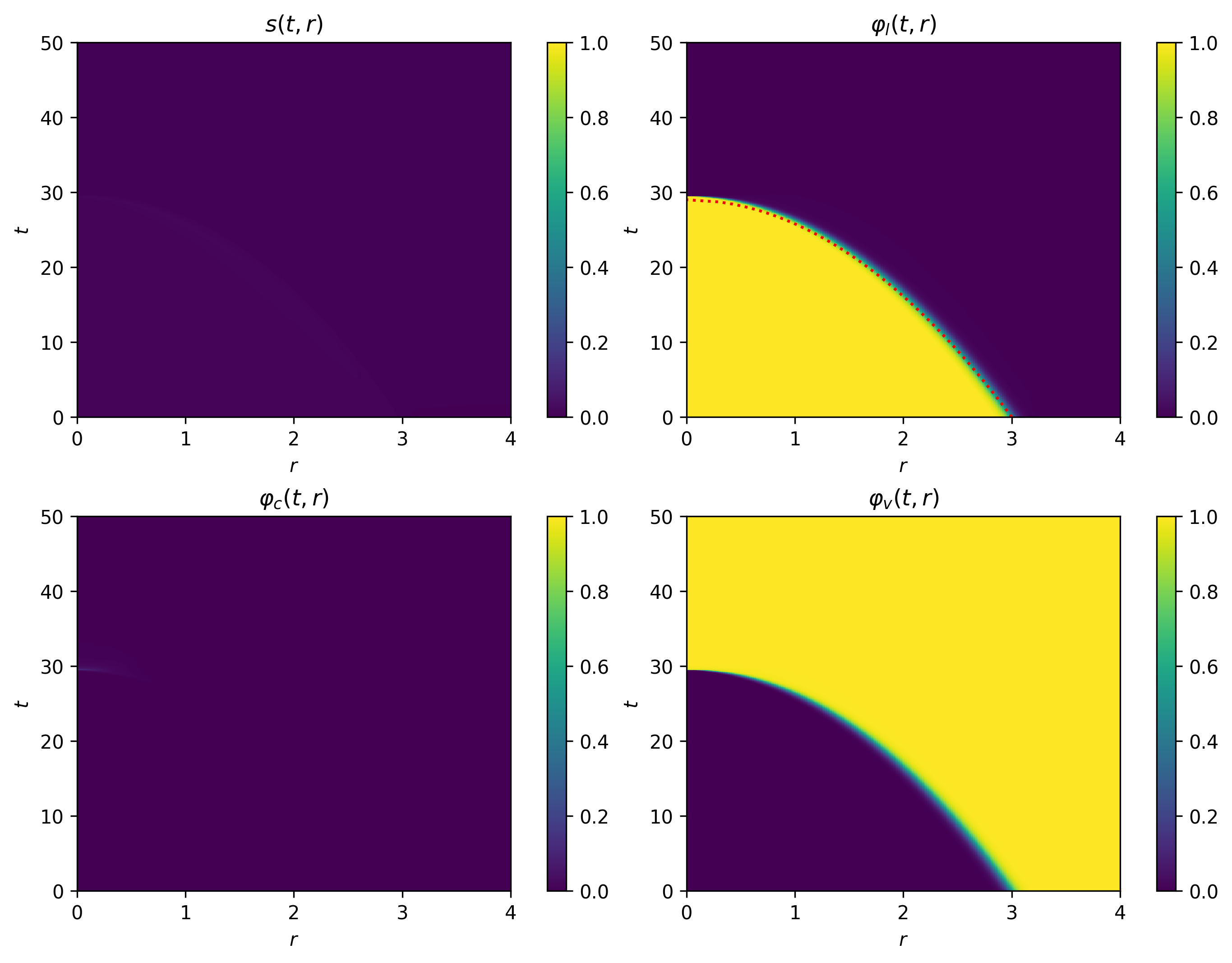}
\caption{Parameters as \Cref{fig:examples_cryst_a} but with $m_{\ell i}=10^{-3}$ for $i\in\{\ell,c,v\},$  where the droplet still completely evaporates. The dotted red curve indicates the function $R(t)=(R_0^{1/\alpha}-Ct)^\alpha$ for $\alpha=0.5,$ as also predicted in \Cref{eq:radius}.}
\label{fig:examples_cryst_a1}
\end{figure}

In the plot for $\varphi_\ell$ in \Cref{fig:examples_cryst_a}  we show the liquid phase field overlaid with the function $R(t)= (R_0^{1/\alpha}-Ct)^\alpha$ and find that $\alpha=0.3$ provides a good fit to the simulation. 
This exponent is close to the expectation $\alpha=1/3$ for the classical Mullins-Sekerka interface law or the canonical droplet dissolution in the LSW theory  \cite{lifshitz1961kinetics} reproduced by Cahn-Hilliard phase-field models \cite{bray2002theory}. 
Furthermore, if we reduce in our model
the mobilities for the liquid phase, \emph{e.g.}, $m_{\ell i} =10^{-3}$ for $i\in\{\ell,c,v\}$ and run the simulation over a time interval $[0,50]$, so that the droplet extinction is dominated by $h_\mathrm{evap}$ 
and diffusion is practically absent, then the radius follows the 
$R(t)$-law with an exponent $\alpha=1/2$, as shown by the red dashed line in \Cref{fig:examples_cryst_a1}.  
This matches the evolution of a droplet boundary by mean curvature flow, which is the sharp-interface limit of the Allen-Cahn equation \cite{Huis84FMCC,Ilma98LMCF}.
The corresponding energy plot in the right panel of  \Cref{fig:energyX_a} shows an almost linear descent of the energy with almost constant dissipation dominated by evaporation $\mathcal{D}^*_\mathrm{evap}$ with the droplet vanishing at $t\approx 29$.

In the corresponding study of \Cref{sec:stagnant} with the one-dimensional ODE sharp-interface model for droplet evaporation in the diffusion-limited regime, also the complete extinction of the droplet can be observed. Equation  \eqref{eq:radius} predicts a law for the evolution of the droplet radius with the exponent $\alpha=1/2$. 
At first glance, this exponent rather matches the evolution of the droplet boundary by mean curvature flow, which is the sharp-interface limit of the Allen-Cahn equation \cite{Huis84FMCC,Ilma98LMCF}.
However, the matching of the exponents in the case $\alpha=1/2$ here is more intricate, as the evolutions in Section \ref{sec:droplet} and \ref{sec-CH} employ slightly different driving forces: The evaporation in the phase-field model is driven primarily by surface tension, giving rise to a mean curvature flow with $\dot{R}(t)\sim -1/R(t)$.  The flux in Equation \eqref{eq:flux} gives rise to $J\sim R$ and thus $\tfrac{\rm d}{\mathrm{d}t}\bigl(\tfrac{4\pi}{3}R^3\bigr)\sim -4\pi R^2/R$, so that also here $\dot{R}\sim -1/R$. Thus, in the considered setting of a spherical droplet, both models give rise to $R^2=R_0^2-Ct$, but due to different driving forces.

\begin{figure}[H]
\centering
\includegraphics[width=0.75\textwidth]{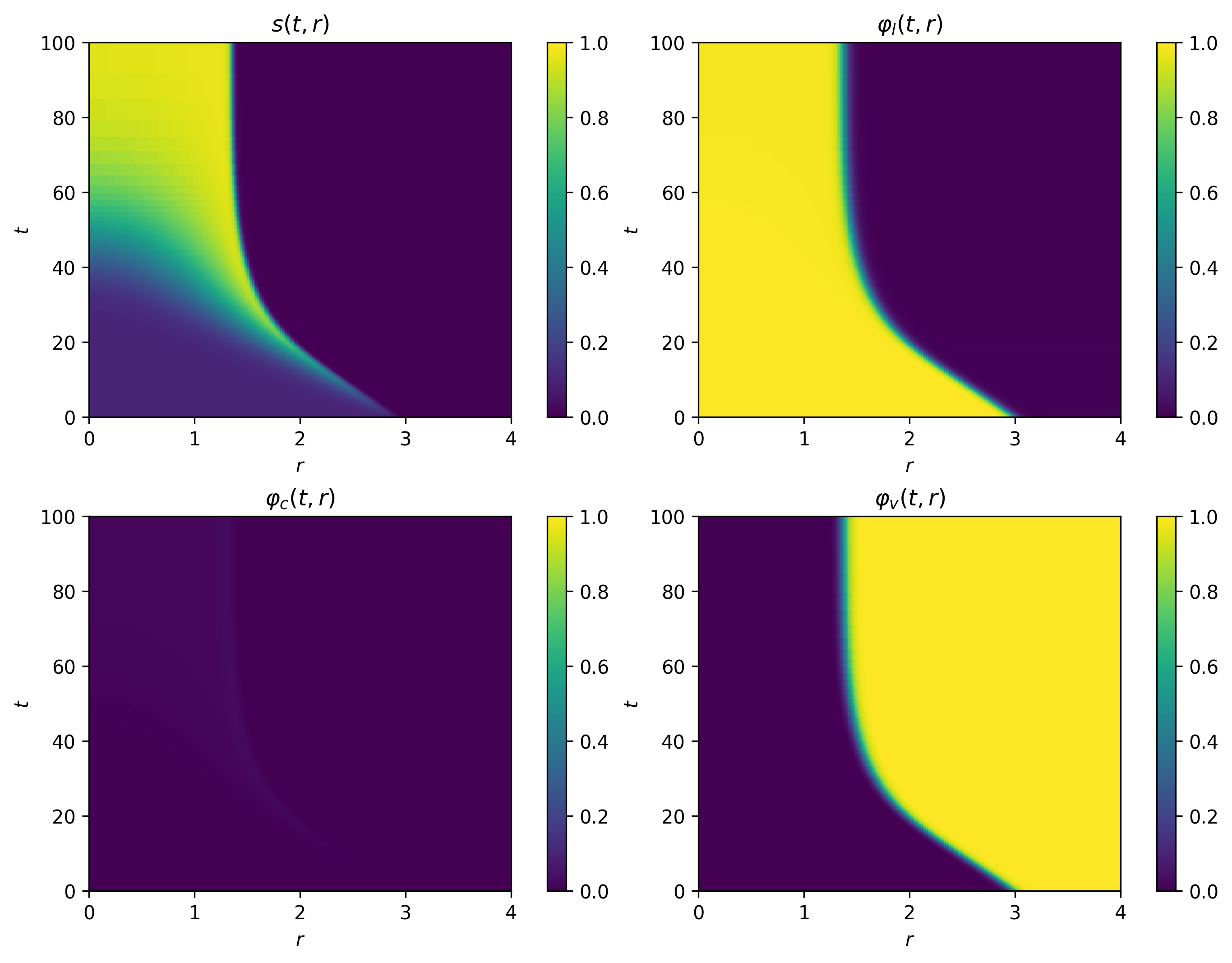}
\caption{Solute concentration $s(t,r)$ and phase fields $\varphi_i(t,r)$ for $i=\{\ell,c,v\}$ as a function of time $t$ and radius $r$ for Example (b) with $\lambda=10$, $\beta=-1$, $\bar{s}^0=10^{-1}$, where the droplet partially evaporates and stabilizes with homogeneous solute concentration.}
\label{fig:examples_cryst_b}
\end{figure}

\begin{figure}[ht]
\centering
\includegraphics[width=0.42\textwidth]{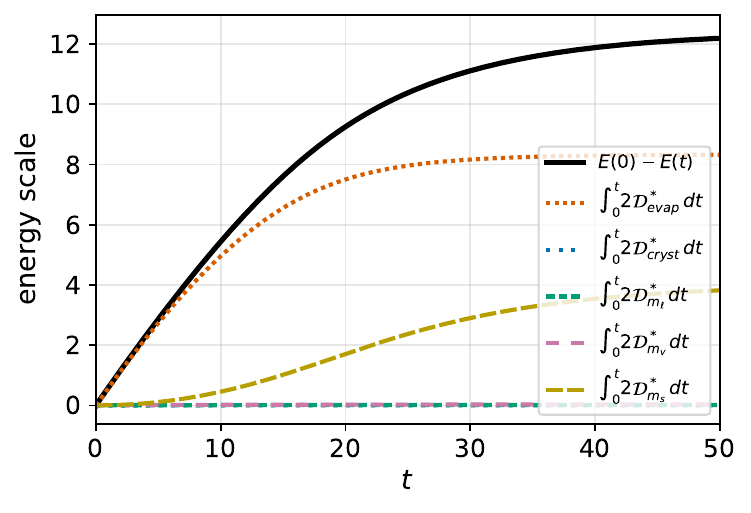}
\caption{Energy $\mathcal{E}$ and different time-integrated contributions to the dissipation $\mathcal{D}^*$ for Example (b) shown in \Cref{fig:examples_cryst_b}.}
\label{fig:energyX_b} 
\end{figure}

\paragraph*{Example (b): Evaporation stops with a solution droplet.} Keeping  $\beta=-1$, but using larger parameters $\lambda$ and $\bar{s}^0,$ \emph{i.e.}, $\lambda=10$ and $\bar{s}^0=0.1,$ results in the droplet evolution depicted in 
\Cref{fig:examples_cryst_b}. Due to evaporation, the droplet size shrinks from its initial radius $R_0=3$ to the radius $R=1.5,$ which is reached at  $t\approx 60$ and then remains constant, cf.\ 
the plot for $\varphi_\ell$. Thanks to the large value of $\lambda$ the solute is confined to the liquid phase, where it diffuses, and, due to the loss in droplet size, the solute concentration is increased over time, cf.\  
the plot for $s$ in \Cref{fig:examples_cryst_b}. As in Example (a), with $\beta=-1,$ also here no crystalline phase is formed, cf.\ 
the plot for $\varphi_c,$ although the solute concentration certainly exceeds the saturation threshold $s_{\mathrm{sat}}=0.3$. Yet, with $\beta=-1,$ it is energetically more favorable to keep $\varphi_c\equiv0$ during the evolution, cf.\ \Cref{choice-Psi}. Hence, again, the vapor phase is given by the complement of the liquid phase, cf.\
the plot of $\varphi_v$ in \Cref{fig:examples_cryst_b}.  
The evolution of the free energy and the time-integrated dissipative contributions during this process is depicted in \Cref{fig:energyX_b}. The energy monotonically decreases until the evaporation process stops at $t\approx 30$ and solute diffusion stops at $t\approx 60$. From then on, also the energy remains constant at a positive value. 

\begin{figure}[H]
\centering
\includegraphics[width=0.75\textwidth]{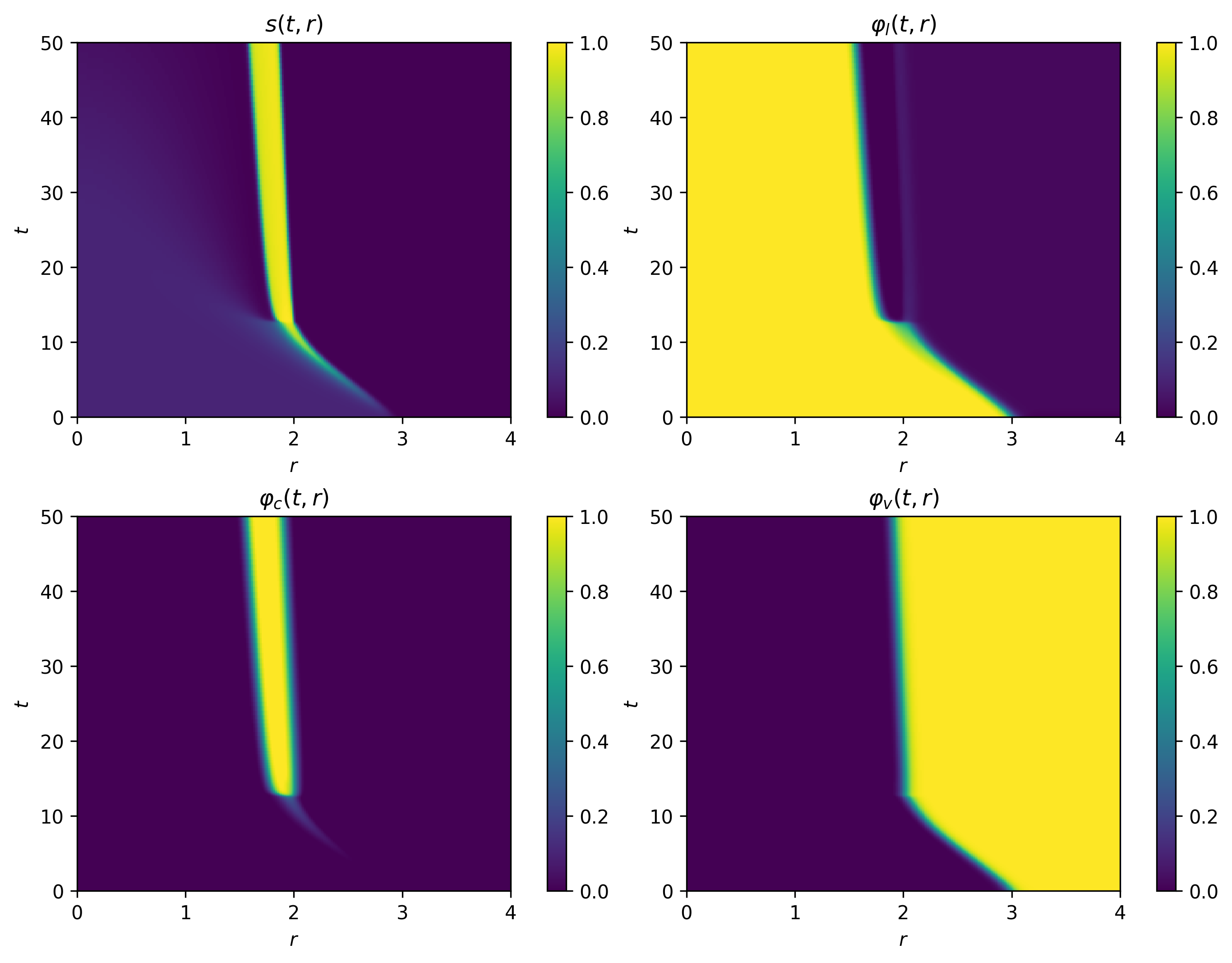}
\caption{Solute concentration $s(t,r)$ and phase fields $\varphi_i(t,r)$ for $i=\{\ell,c,v\}$ as a function of time $t$ and radius $r$ for Example (c) with $\lambda=10$, $\beta=-10$, $\bar{s}^0=10^{-1}$, where the droplet evaporates and forms a crystalline crust.}
\label{fig:examples_cryst_c}
\end{figure}

\begin{figure}[ht]
\centering
\includegraphics[width=0.42\textwidth]{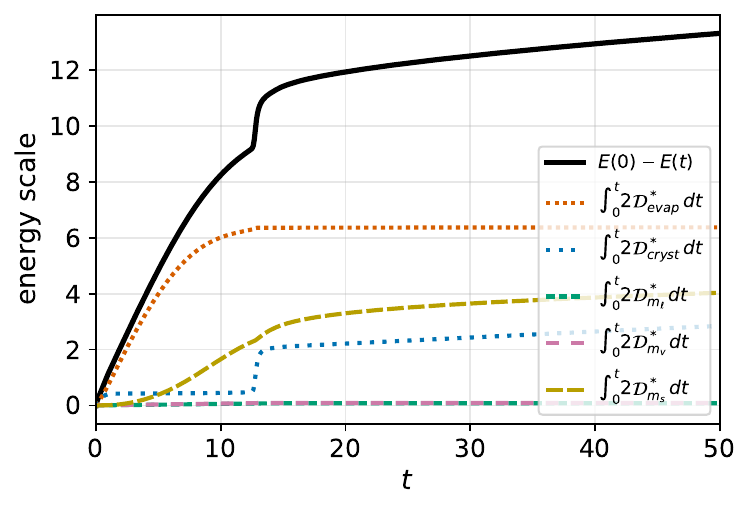}
\caption{Energy $\mathcal{E}$ and different time-integrated contributions to the dissipation $\mathcal{D}^*$ for Example (c) shown in \Cref{fig:examples_cryst_c}.}
\label{fig:energyX_c} 
\end{figure}

\paragraph*{Example (c): Droplet with crystalline crust.} 
Here we keep $\lambda=10$ and $\bar{s}^0=0.1$ as in Example (b), but additionally decrease $\beta$ to $\beta=-10,$ which now favors the creation of a crystalline phase. The simulation results are depicted in \Cref{fig:examples_cryst_c}. 
As can be seen in the plot of $\varphi_\ell,$ evaporation first decreases the droplet radius from initially $R_0=3$ to $R=2.5$ at time $t=5$. Then, additionally also the crystallization process sets in. The crystalline phase forms at the interface between liquid and vapor, as is favored by $h_{\mathrm{cryst}}(\bs{q})$ 
in \Cref{choice-h} and also by the values of the surface tensions, cf.\ \Cref{surftens} and \Cref{tab:params}. Observe that the latter also allow for a liquid layer between the crystal and the vapor phase, as can be detected in the
plot of $\varphi_\ell$ 
from $t\approx 13$ on. Then, also the crystallization process rapidly increases. Since the crystal phase is immobile, \emph{i.e.},  $m_c=0$ in \Cref{choice-m}, it thus creates a crust at the droplet surface. Comparing the 
plots of $s$ and $\varphi_c$ in \Cref{fig:examples_cryst_c}, 
one can see that this sudden increase of the crystal phase goes along with a high amount of solute significantly exceeding the saturation threshold. Due to the formation of a crust at the droplet surface, the evaporation process is significantly slowed down after $t\approx 13,$ but does not come to a halt.  This behavior can also be confirmed by the evolution of the free energy and time-integrated dissipative contributions over time as depicted in \Cref{fig:energyX_c}. Here one sees a first decrease in energy due to evaporation, clearly followed by a sudden energy drop due to a rapid formation of a crystalline crust at $t\approx 13$, which then continues with a moderate energy decrease due to the solute diffusion. 

\begin{figure}[H]
\centering
\includegraphics[width=0.75\textwidth]{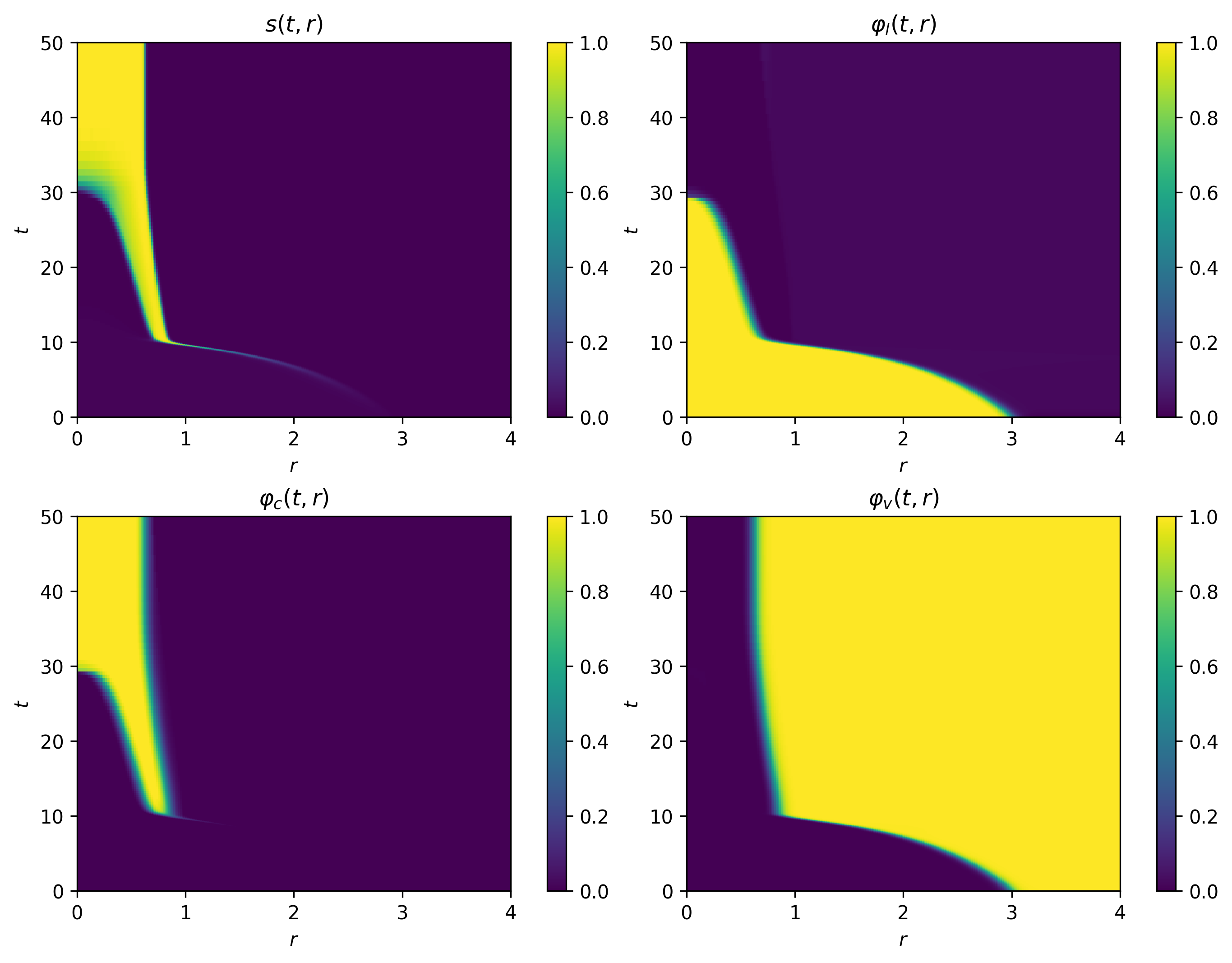}
\caption{Solute concentration $s(t,r)$ and phase fields $\varphi_i(t,r)$ for $i=\{\ell,c,v\}$ as a function of time $t$ and radius $r$ for Example (d) with $\lambda=10$, $\beta=-10$, $\bar{s}^0=10^{-2}$, where the droplet evaporates, reaches $s=1,$ and then precipitates with liquid phase vanishing.}
\label{fig:examples_cryst_d}
\end{figure}

\begin{figure}[ht]
\centering
\includegraphics[width=0.42\textwidth]{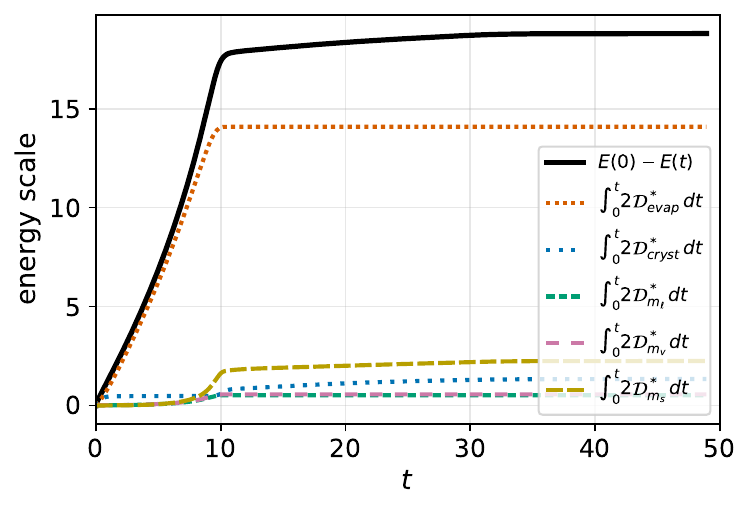}

\caption{Energy $\mathcal{E}$ and different time-integrated contributions to the dissipation $\mathcal{D}^*$ for Example (d) shown in \Cref{fig:examples_cryst_d}.}
\label{fig:energyX_d} 
\end{figure}

\paragraph*{Example (d): Droplet evaporating and forming a crystal.}
Here we again choose $\lambda=10,$ $\beta=-10,$ but a low solute concentration $\bar s^0=10^{-2}$. The low value of $\beta$ again favors the formation of a crystal phase. However, due to the lower solute concentration the crystal phase forms more slowly, 
cf.\ plot of $\varphi_c$ in  \Cref{fig:examples_cryst_d}, 
and evaporation progresses, first quickly, till a radius of $R\approx 0.8$ is reached at $t\approx 11,$ cf.\ 
the plot of $\varphi_\ell$. Then it slows down, but progresses till complete evaporation of the liquid phase is reached at $t=30$. The slowing-down of the evaporation process  is due to a speed-up of crystallization, which first primarily takes place at the interface between the liquid and the vapor phase. The solute is confined to the crystal and the liquid phase, but due to the lower initial concentration, it can just slow down the evaporation process, but not bring it to a halt. Therefore, evaporation continues, so that at $t=30$ a crystal is left. The  profiles of the free energy and the time-integrated dissipative contributions depicted in \Cref{fig:energyX_d}  also show a rapid decrease of the free energy due to the fast evaporation till $t\approx 11,$ followed by a slow energy decrease due to crystallization and slow evaporation.  The energy reaches the value of approximately zero at $t\approx 30,$ when the droplet has fully crystallized and the liquid phase has disappeared. 

\section{Conclusions and Outlook  }\label{sec-conclusion}

In this paper we treat three different aspects of the dynamics of aerosol-size distributions  on different length and time scales.

In \Cref{sec-sed-model}, we  present the theoretical framework to describe the coupled effects of evaporation and sedimentation on the dynamics of aerosol droplet ensembles in terms of their size distribution, which allows us to
calculate the fraction of virions that remain suspended in air as a function of
time and relative humidity. An exact solution of the underlying population
dynamics equation is derived, which can be evaluated numerically for
arbitrary initial droplet distributions. The results show that the droplet size
distribution has a significant effect on the fraction of virions that remain
suspended in air. 

In \Cref{sec:reflection}, we  employ Molecular Dynamics simulations to
determine the molecular reflection coefficient of single water molecules that encounter the
vapor-liquid water interface from the vapor phase
as a function of the angle and velocity of the impinging water molecule. 
The molecular reflection coefficient is a key input parameter for the calculation of the water condensation rate in \Cref{sec-sed-model} and has in all previous work been set to zero. Our Molecular Dynamics simulations indeed 
 show that the reflection coefficient is
negligibly small for water molecules impinging onto the liquid phase with velocities
typical for room temperature, but can become significant for larger velocities
and large angles of incidence.

Finally, in \Cref{sec-CH} we derive a thermodynamically consistent three-phase diffuse-interface model in terms of a coupled Cahn-Hilliard/Allen-Cahn model, featuring a liquid, a vapor, and a crystalline phase, where in the liquid a solute species is present  that  may crystallize. We discuss the gradient-flow structure of the model and provide a weak formulation. Based on this, we introduce a discretization in space and time, and carry out numerical simulations for a broad range of experimentally relevant droplet-drying scenarios. In this way, we show that the diffuse-interface model is able to capture features that are also observed with the one-dimensional model of \Cref{sec-sed-model} and to generalize it to the process of crystallization and crust formation. 

While a mathematically rigorous multiscale approach is beyond the scope of this paper, the three different aspects treated in the preceding Sections highlight the importance of a scale-bridging approach for describing the evaporation and sedimentation of aerosol ensembles.
We believe that in the future, machine-learning approaches will be able to systematically match energetic and dissipative parameters  in the phase-field model to constitutive molecular kinetics laws in order to capture the drying dynamics more realistically.

\begin{figure}[ht!]
\includegraphics[width=0.33\textwidth]{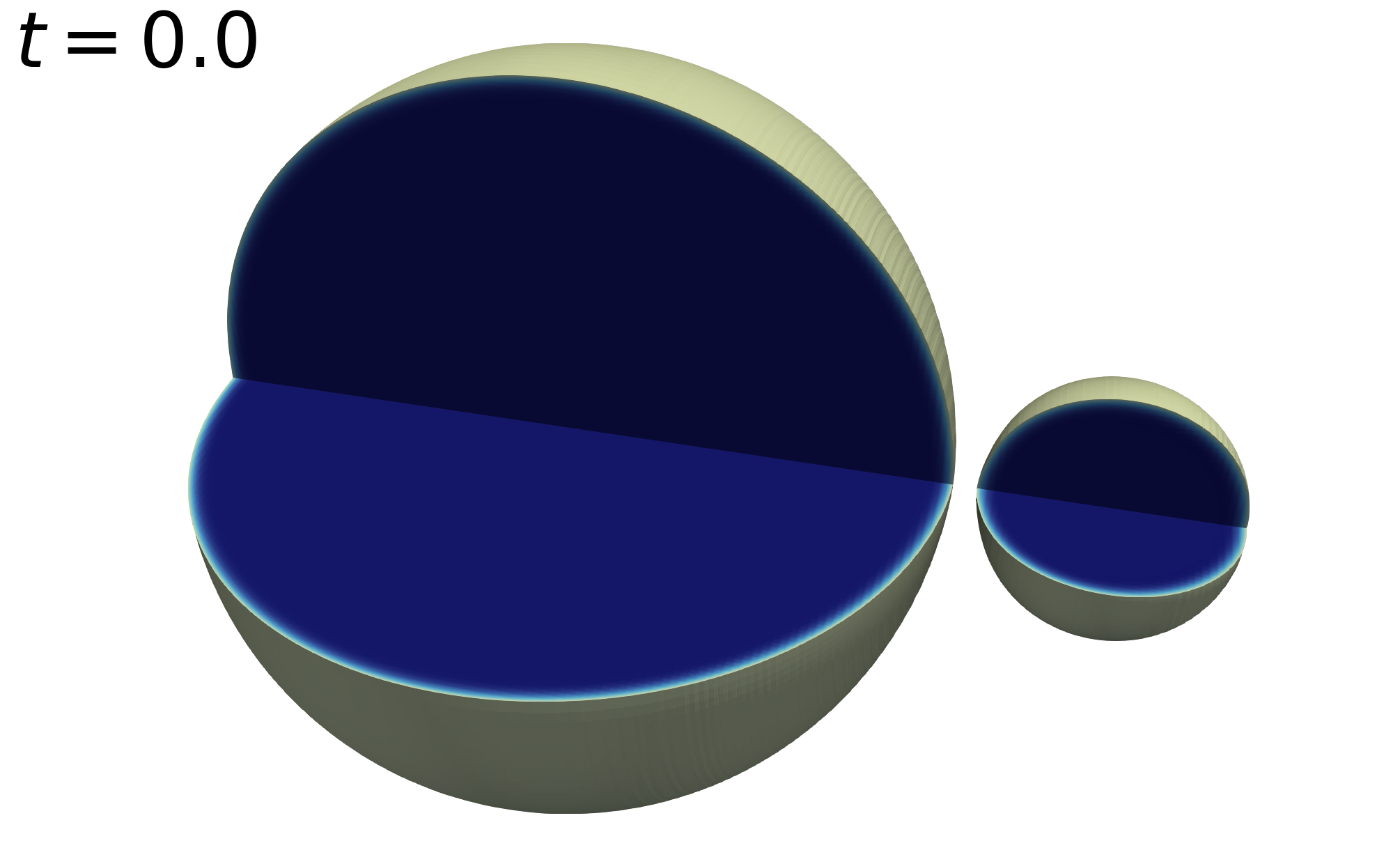}%
\includegraphics[width=0.33\textwidth]{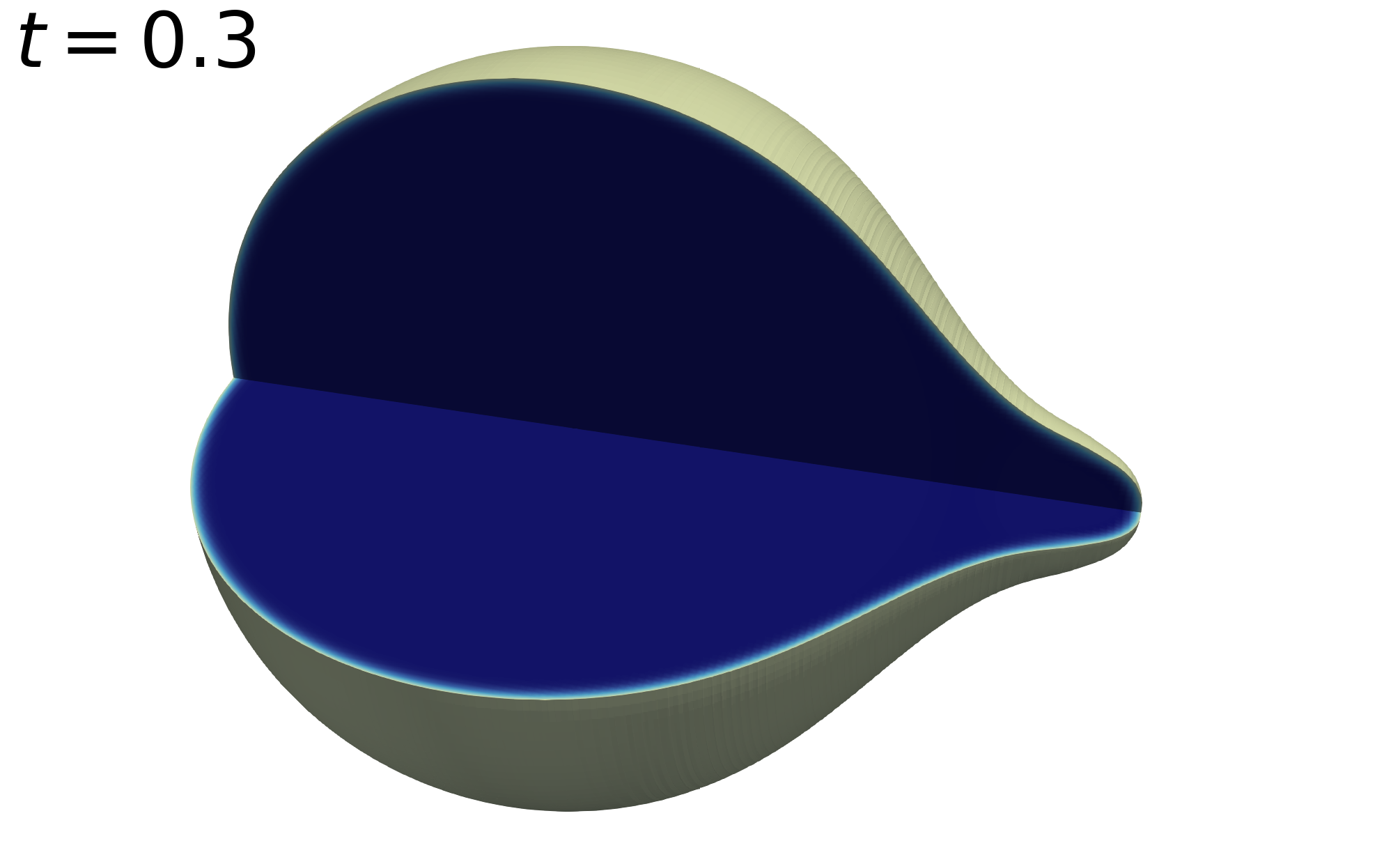}%
\includegraphics[width=0.33\textwidth]{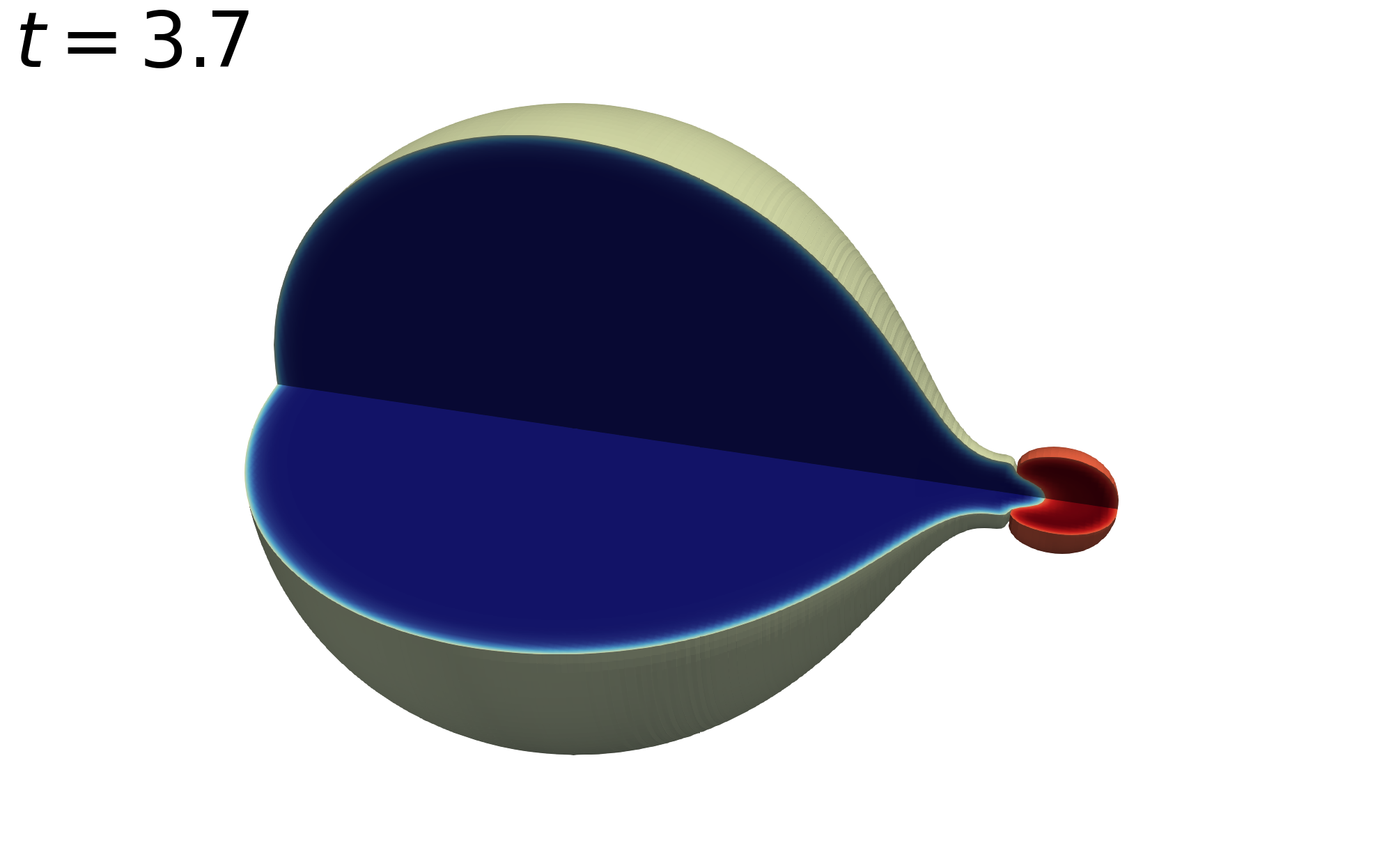}%

\includegraphics[width=0.33\textwidth]{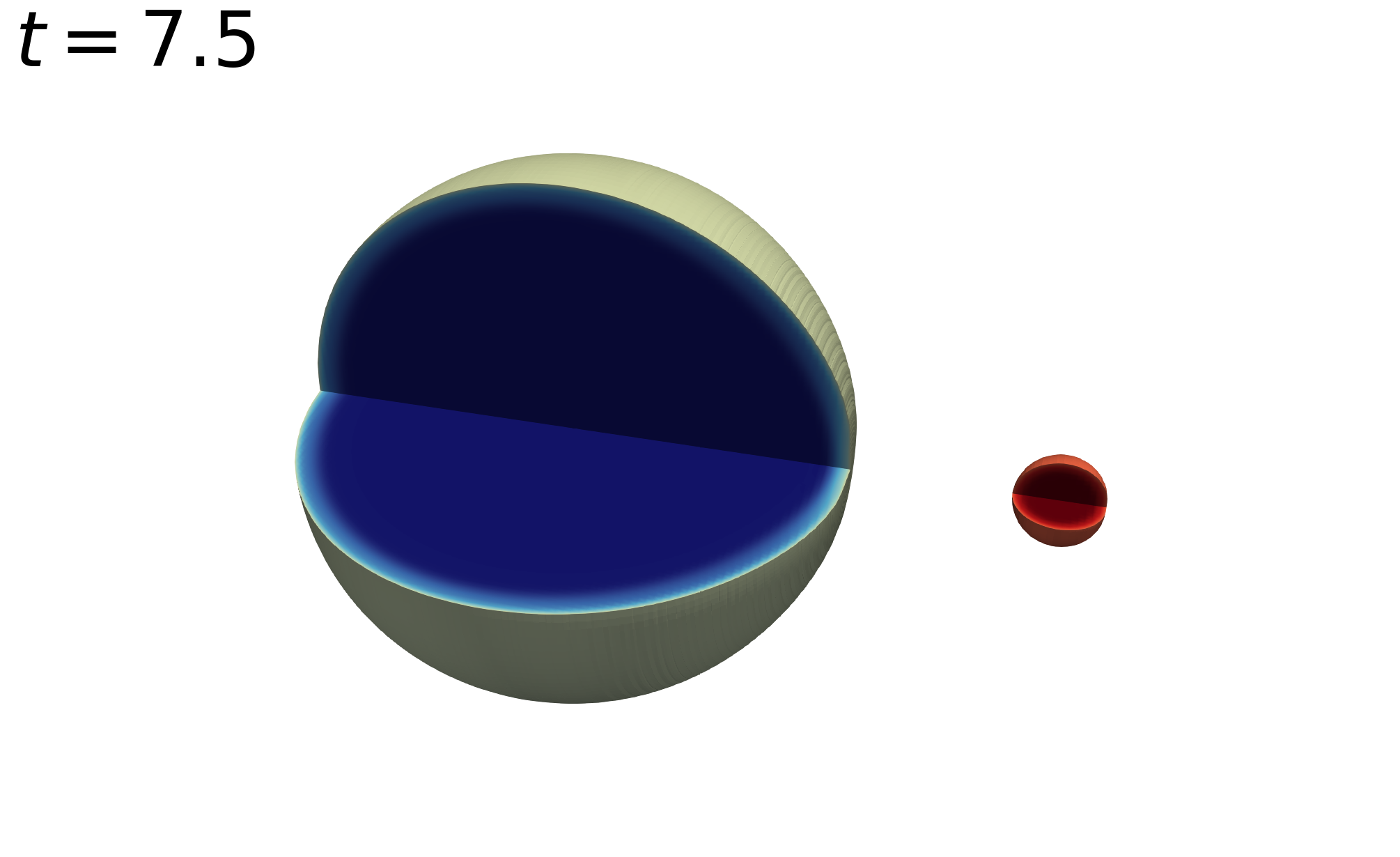}%
\includegraphics[width=0.33\textwidth]{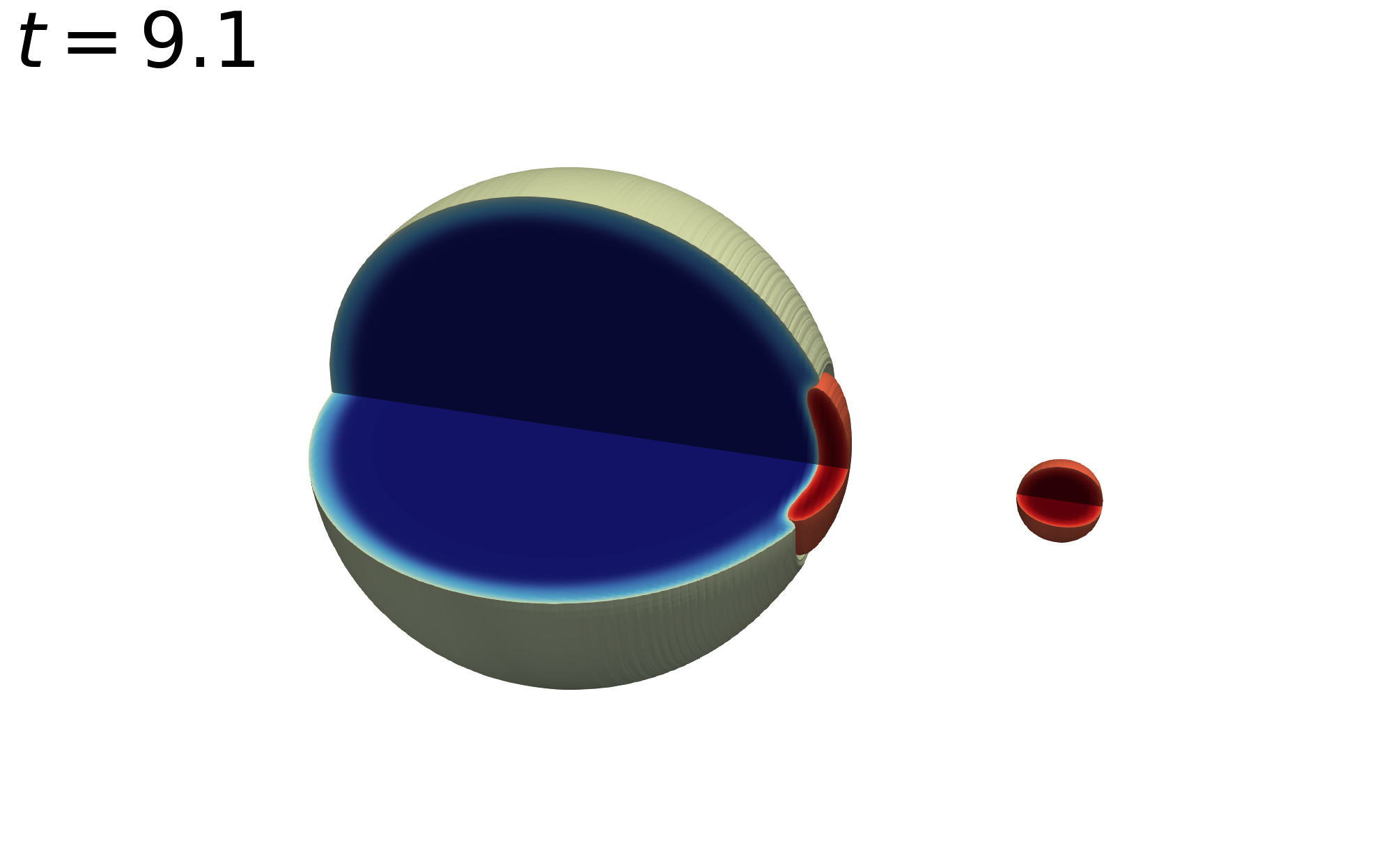}%
\includegraphics[width=0.33\textwidth]{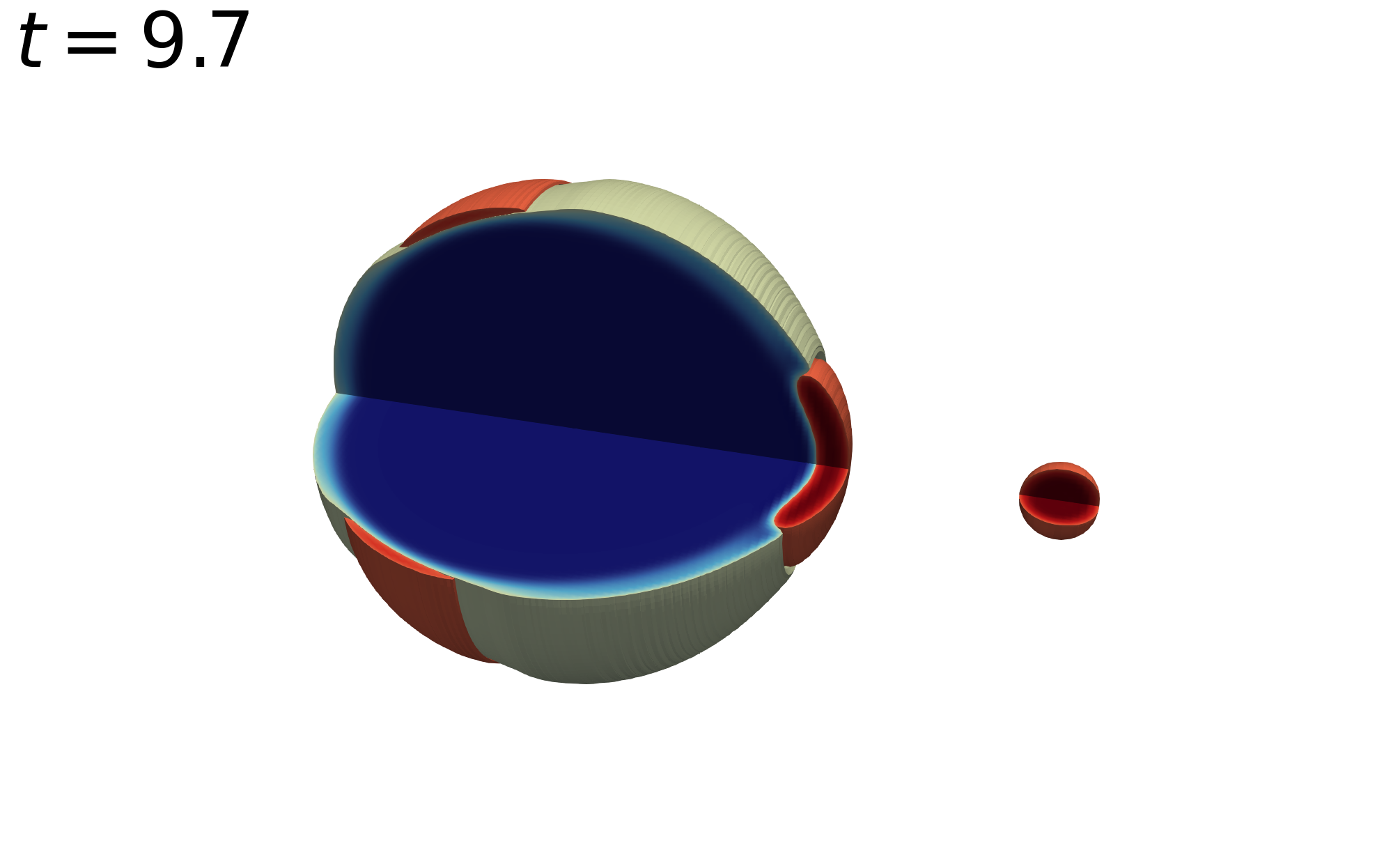}%

\includegraphics[width=0.33\textwidth]{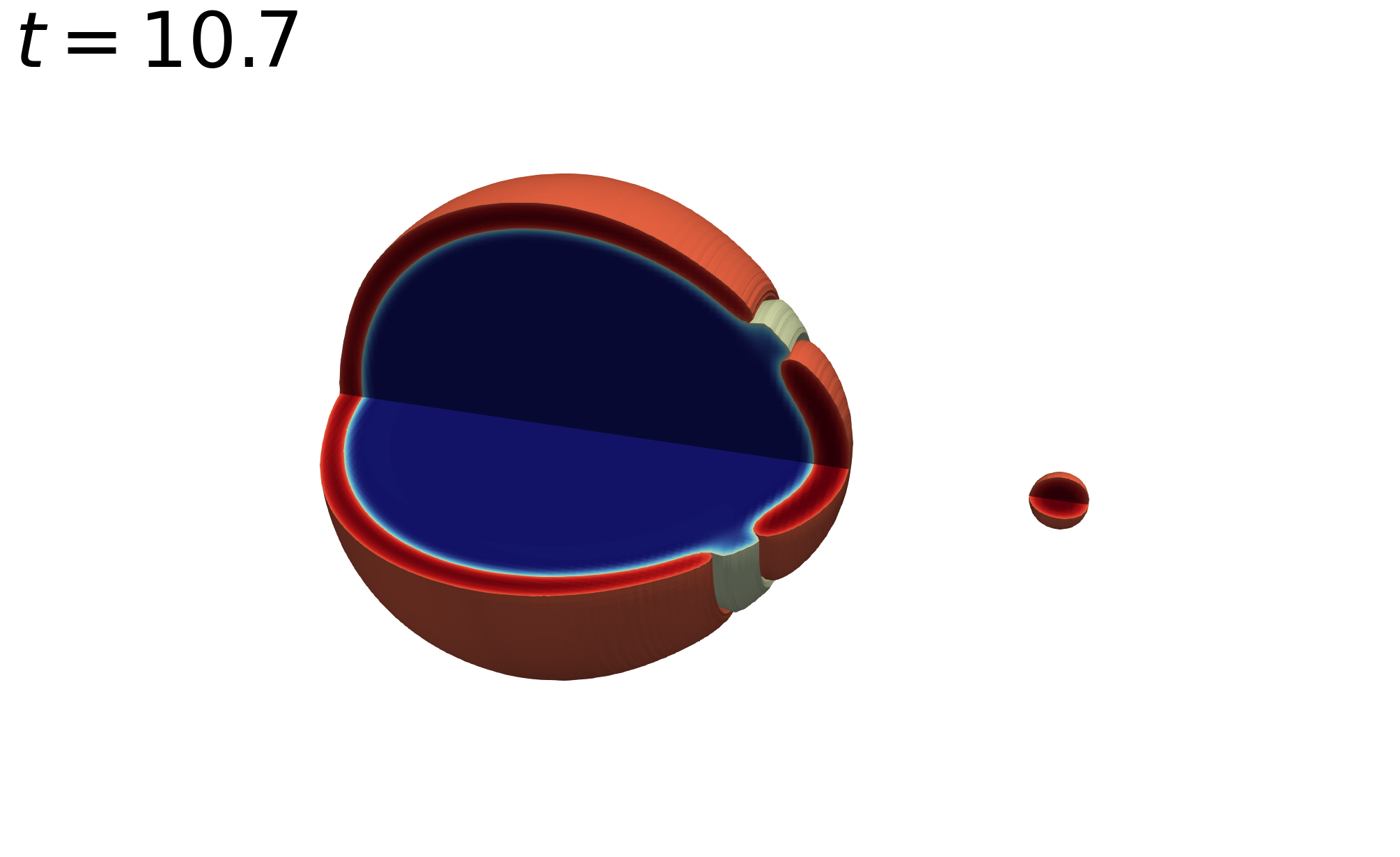}%
\includegraphics[width=0.33\textwidth]{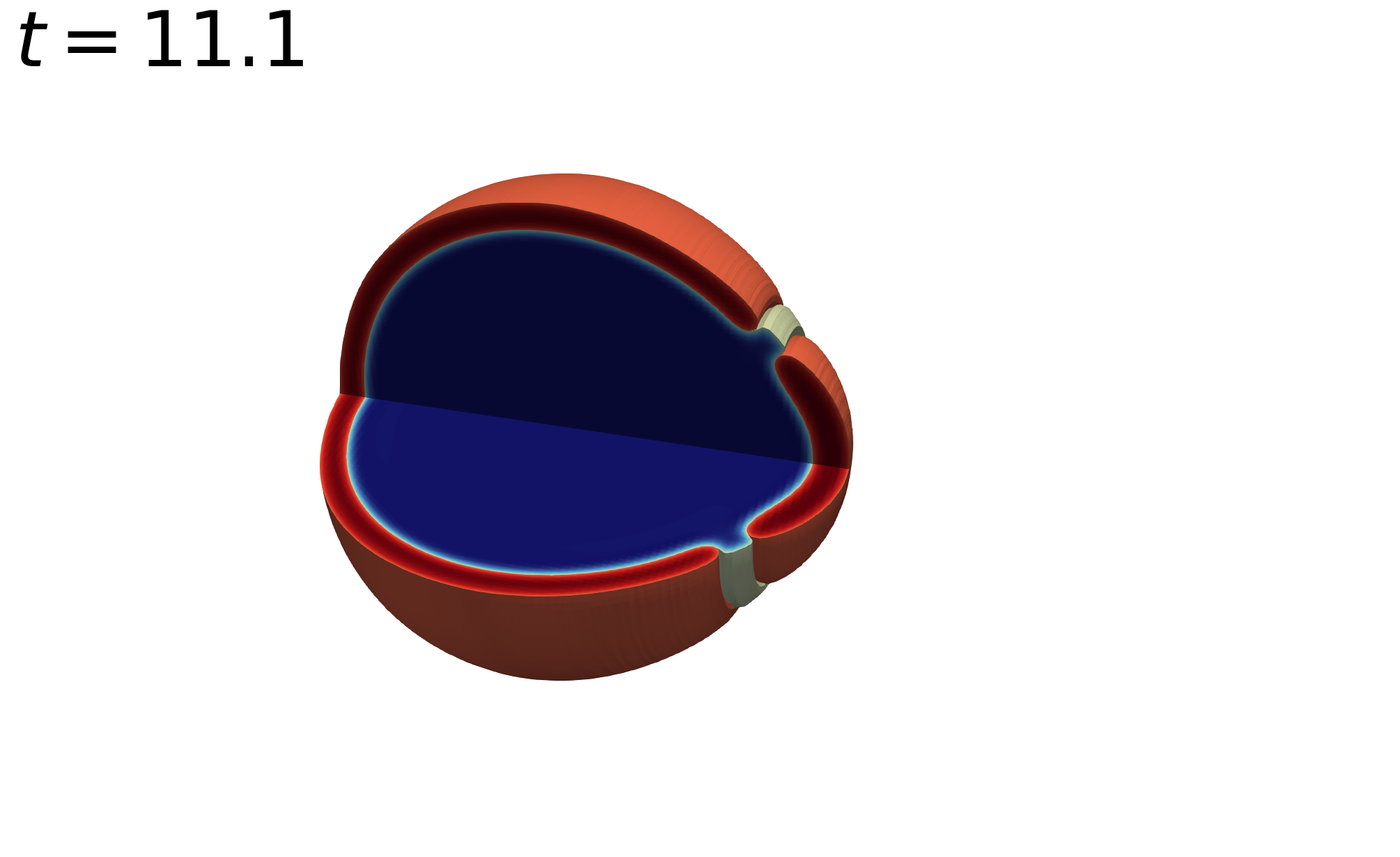}%
\includegraphics[width=0.33\textwidth]{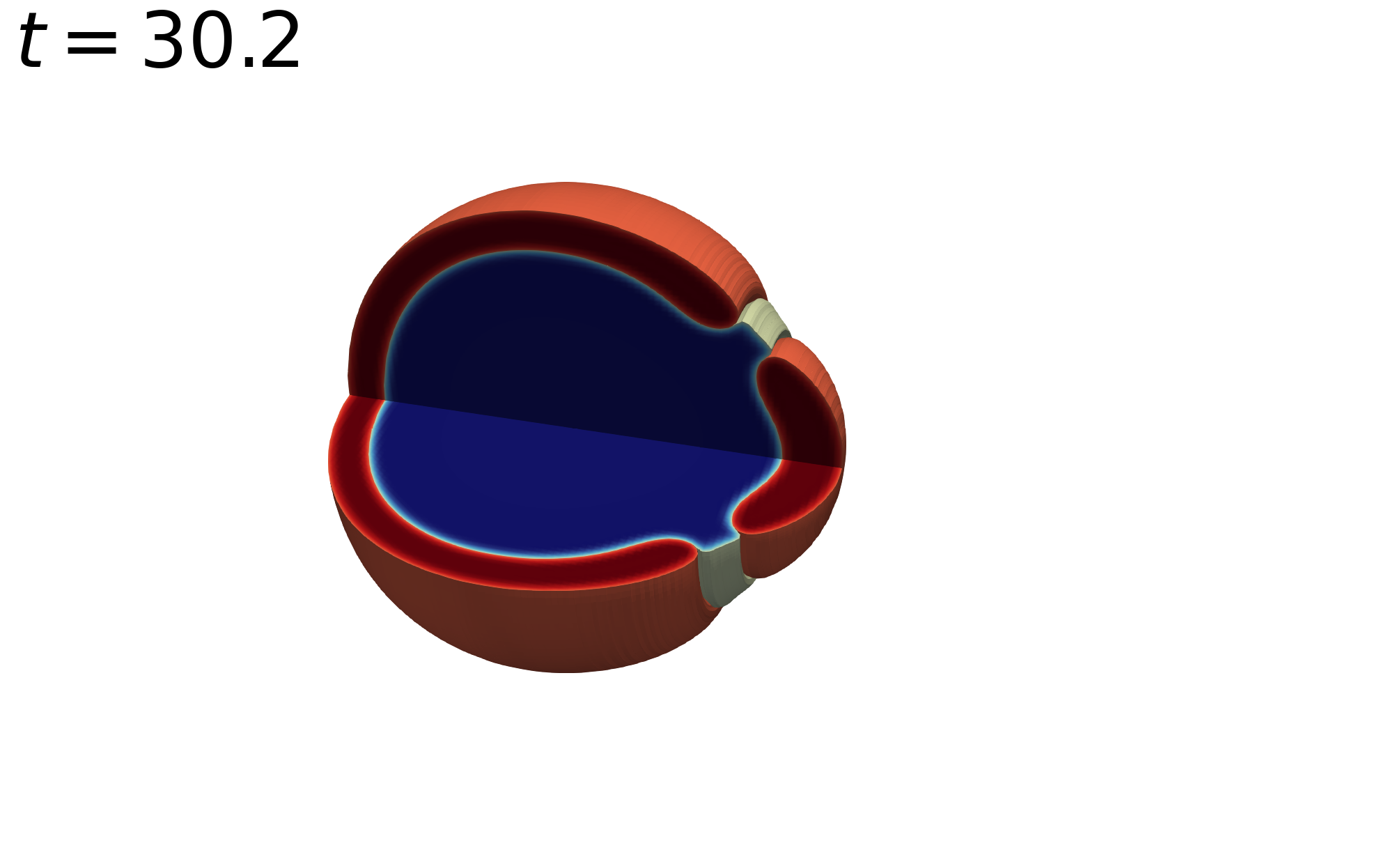}
\caption{Axisymmetric solution for case (c) at different times $t$, shown as a cutaway three-dimensional reconstruction obtained by rotating the two-dimensional $(r,z)$ solution about the symmetry axis. The blue shaded region indicates the liquid phase $\varphi_\ell \ge1/2$, while the red region indicates the crystalline phase $\varphi_c \ge 1/2$.}
\label{fig:3ddroplet}
\end{figure}

\begin{figure}[h]
\mbox{\includegraphics[height=0.26\textwidth,trim=13.5cm 0 30cm 0,clip]{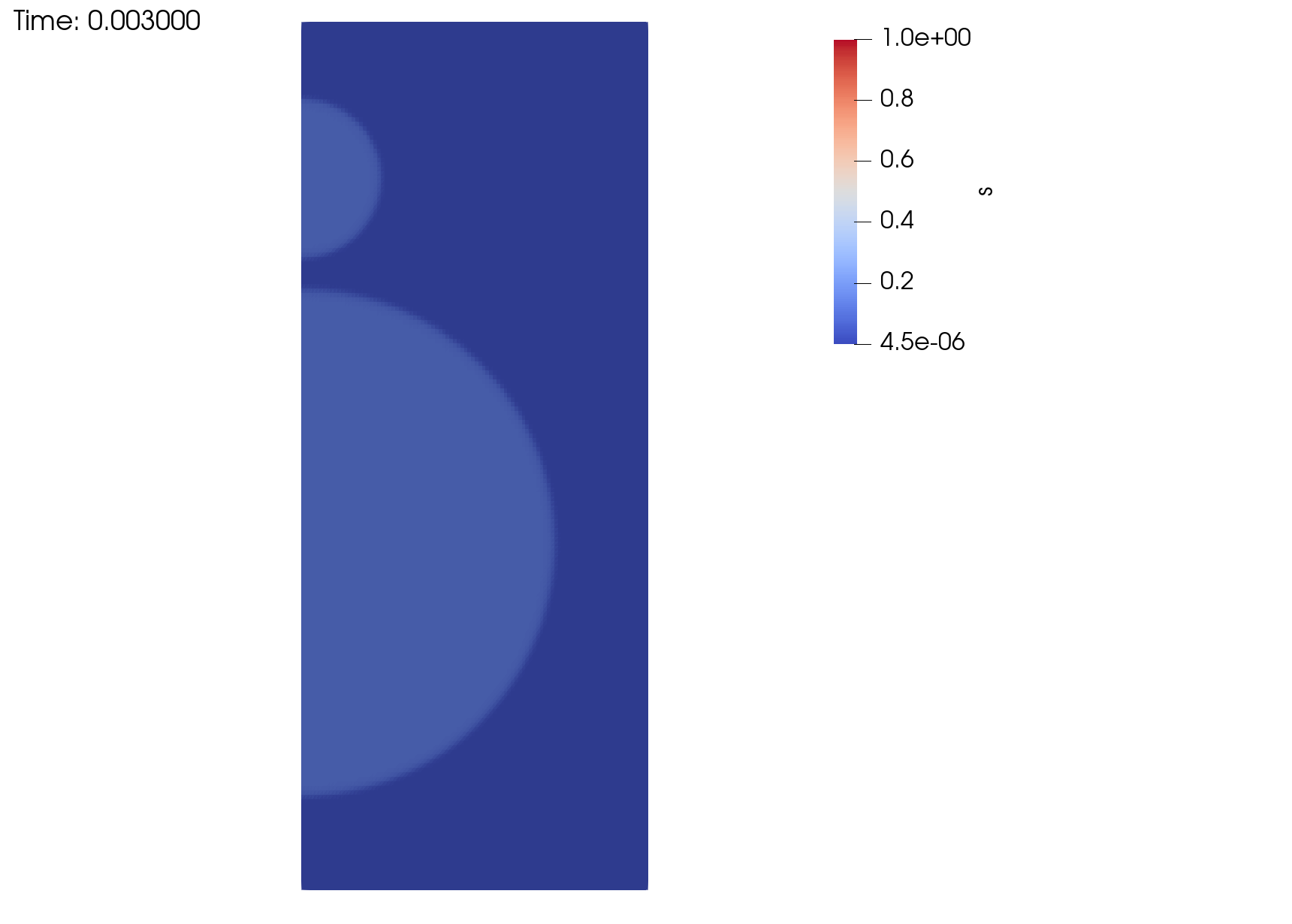}}\hfill%
\mbox{\includegraphics[height=0.26\textwidth,trim=13.5cm 0 30cm 0,clip]{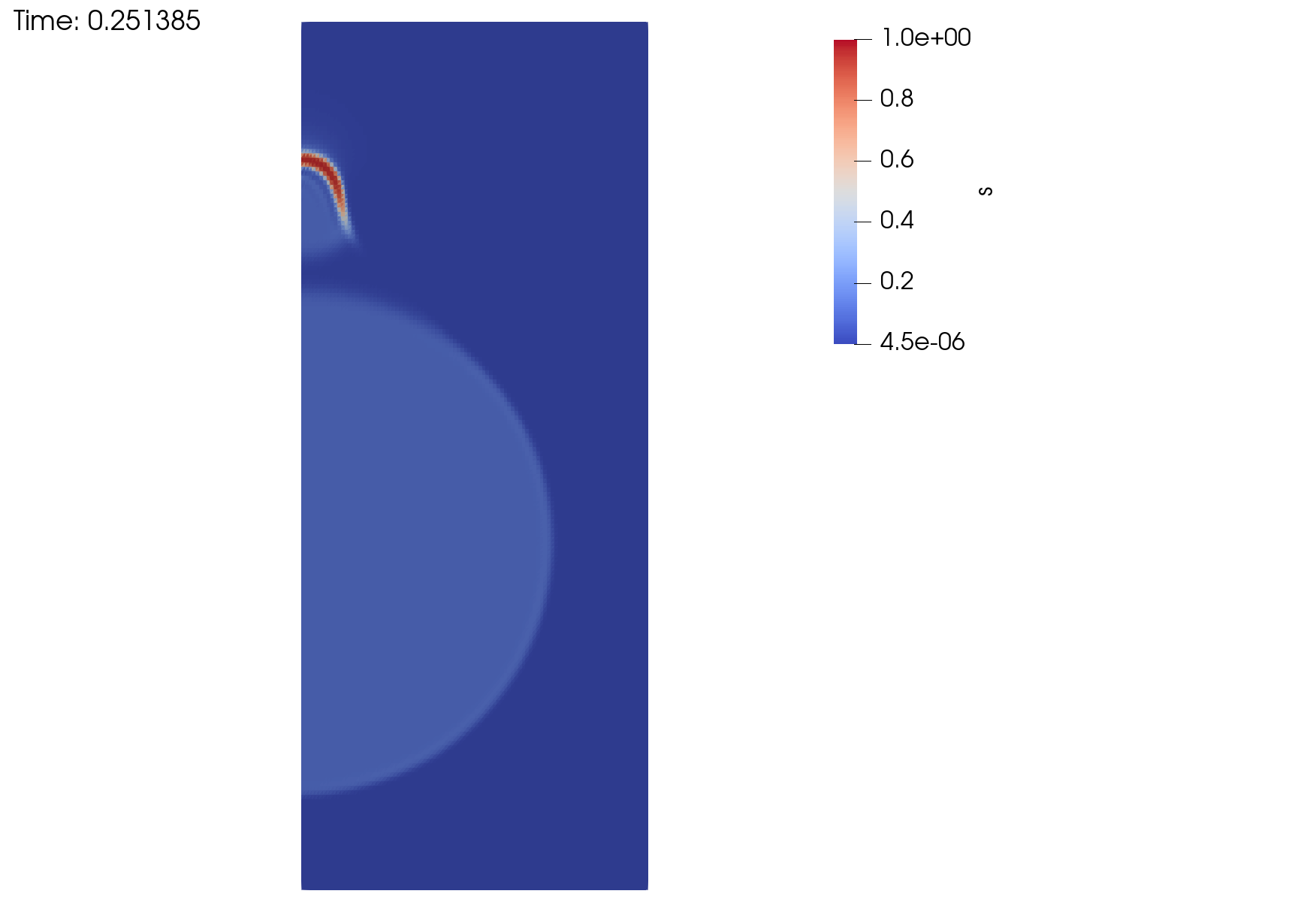}}\hfill%
\mbox{\includegraphics[height=0.26\textwidth,trim=13.5cm 0 30cm 0,clip]{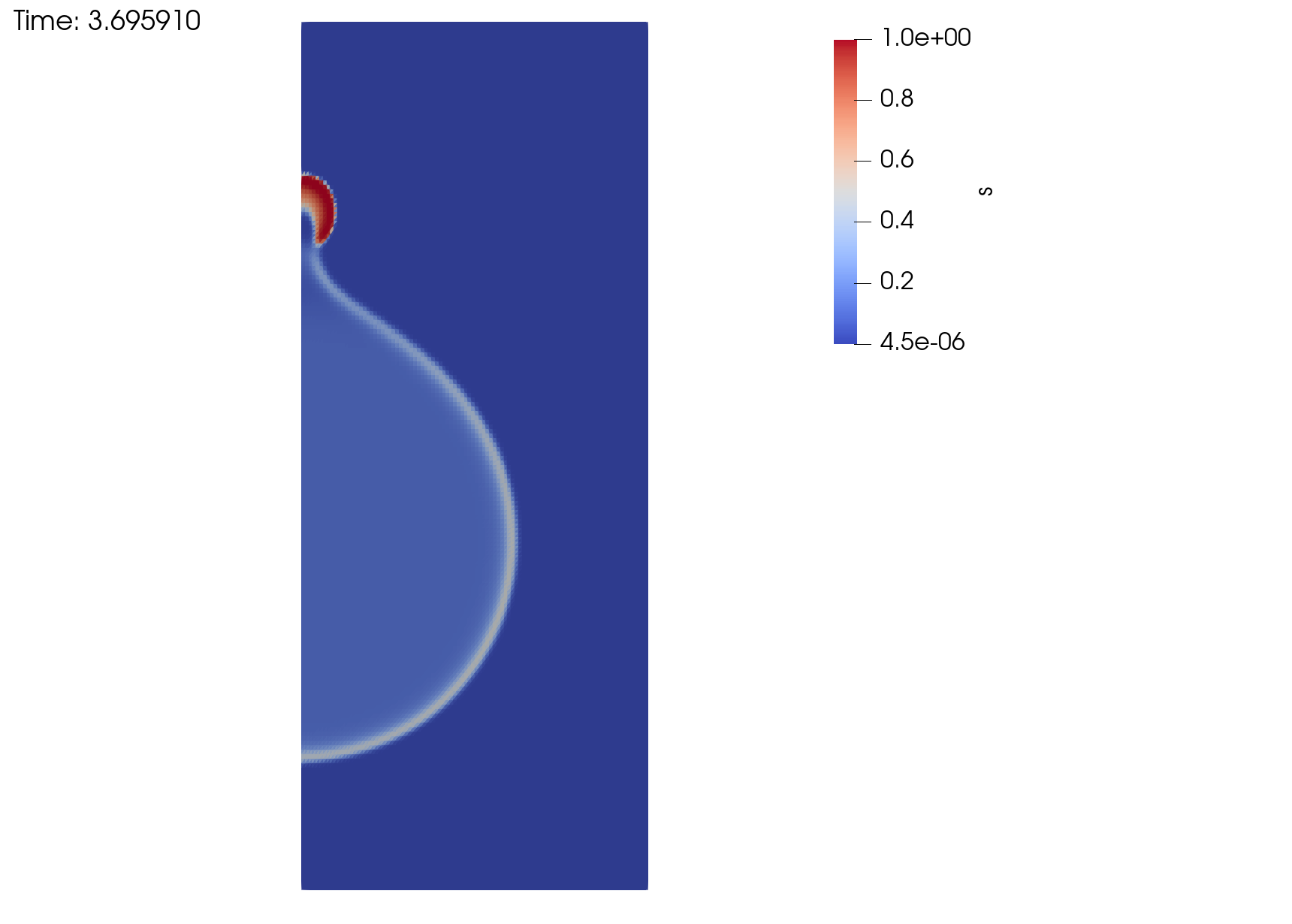}}\hfill%
\mbox{\includegraphics[height=0.26\textwidth,trim=13.5cm 0 30cm 0,clip]{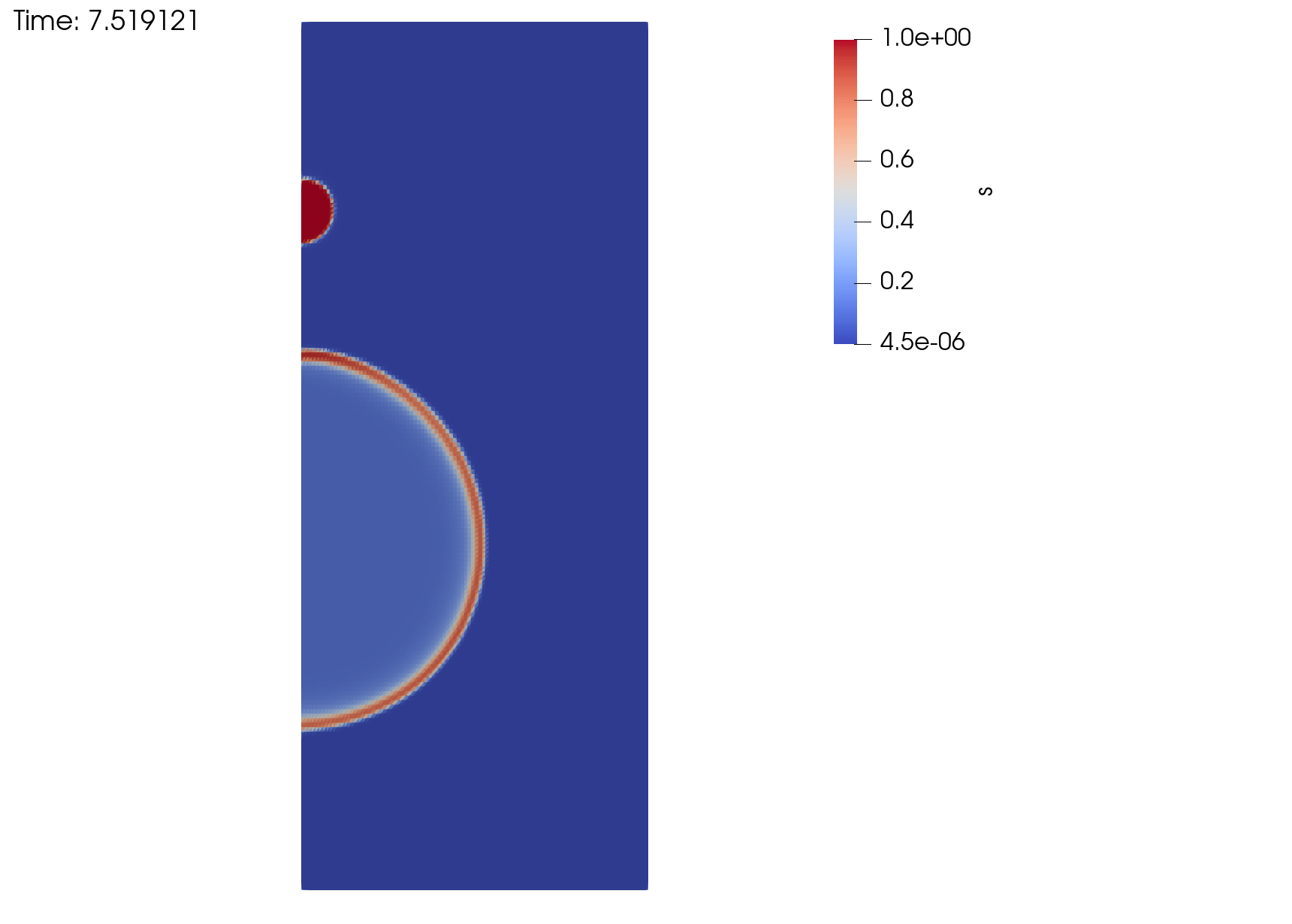}}\hfill%
\mbox{\includegraphics[height=0.26\textwidth,trim=13.5cm 0 30cm 0,clip]{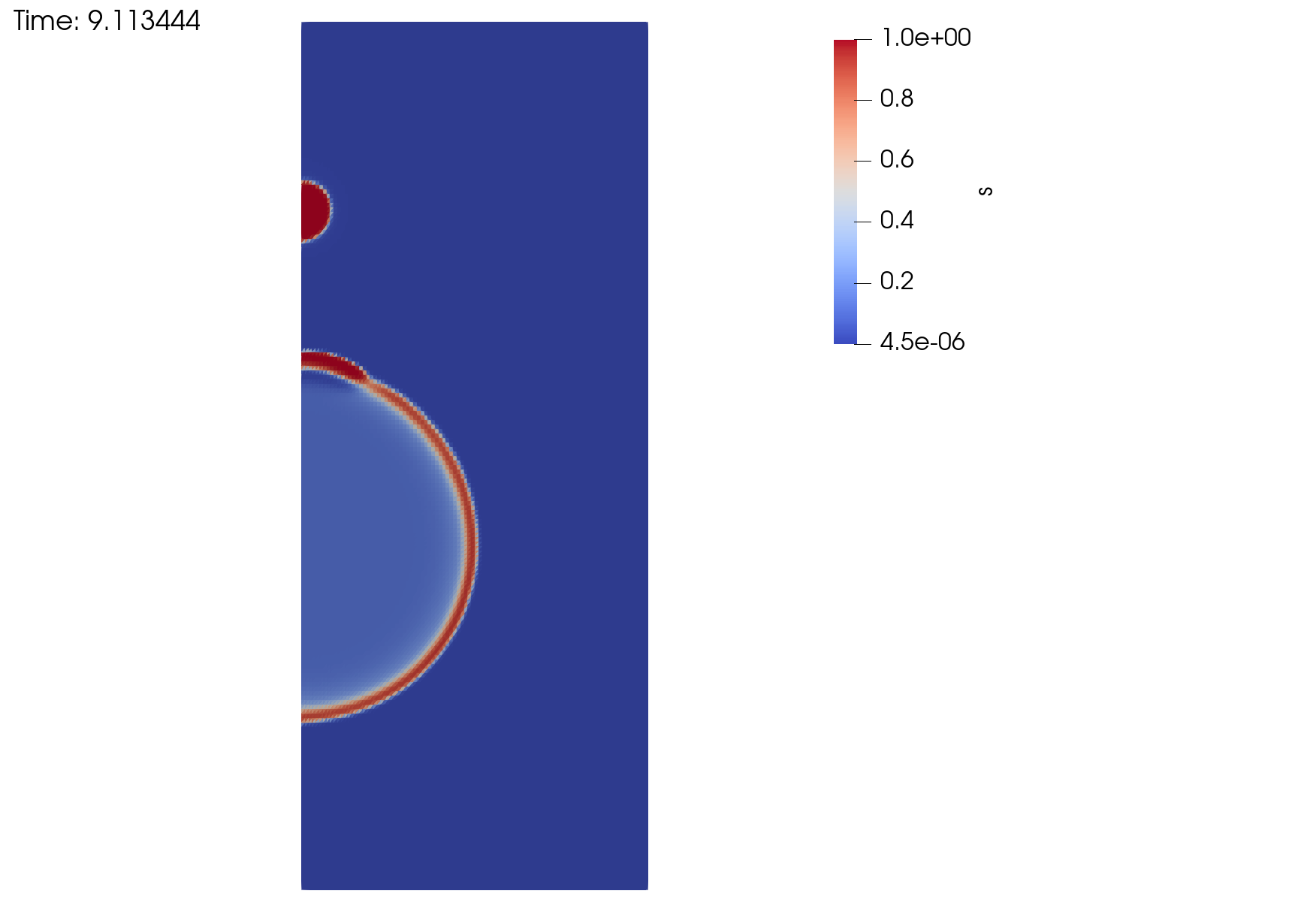}}\hfill%
\mbox{\includegraphics[height=0.26\textwidth,trim=13.5cm 0 30cm 0,clip]{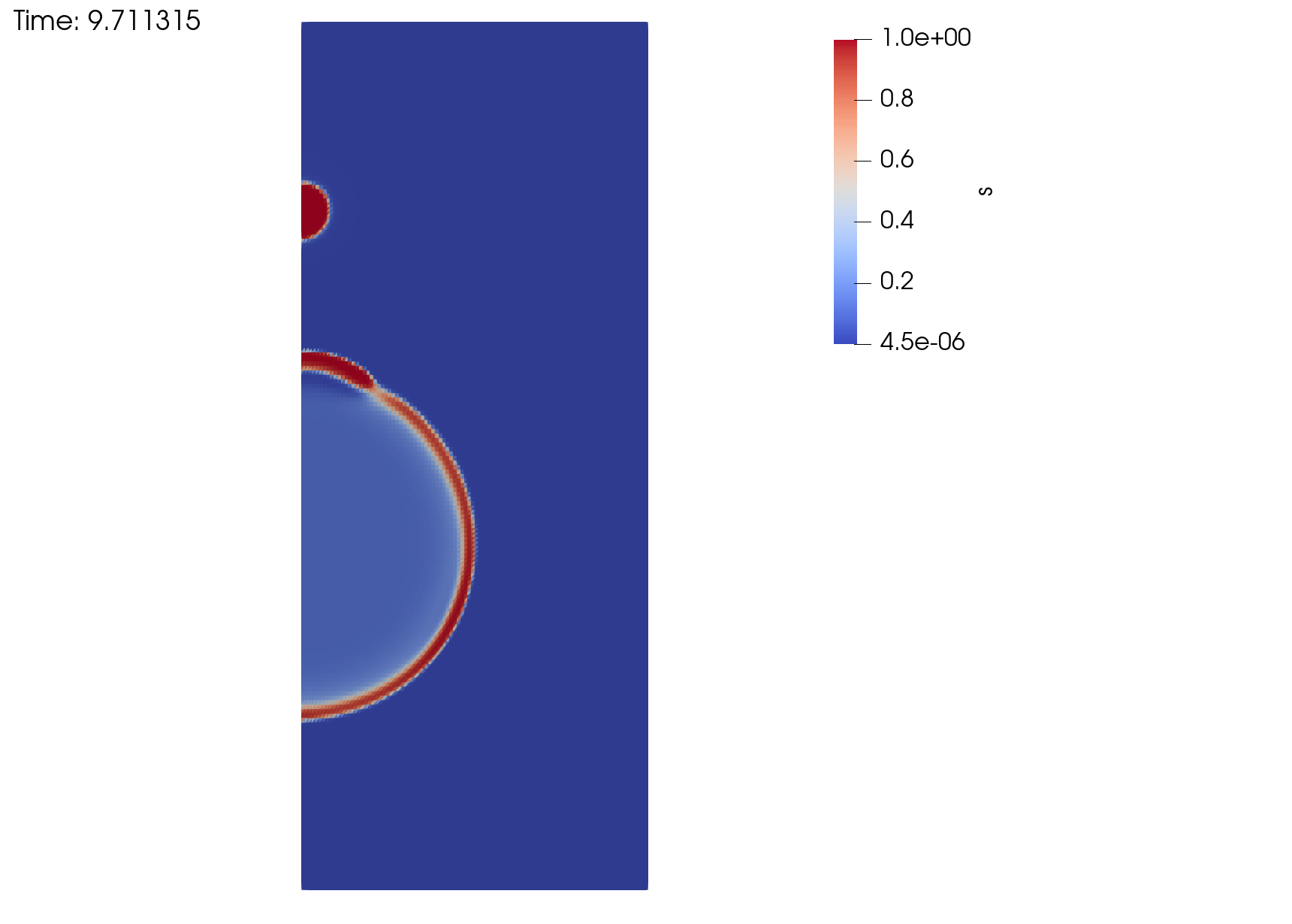}}\hfill%
\mbox{\includegraphics[height=0.26\textwidth,trim=13.5cm 0 30cm 0,clip]{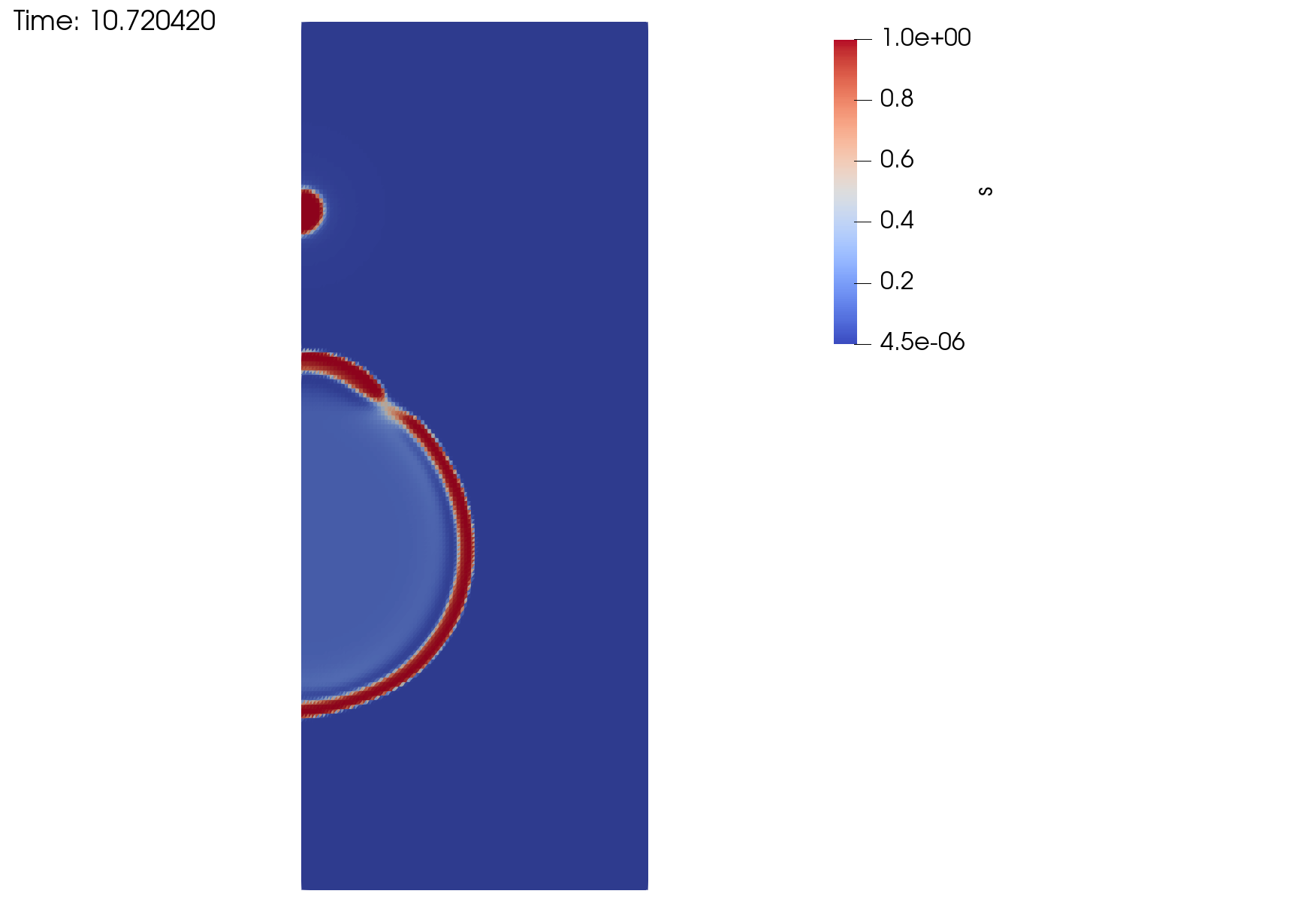}}\hfill%
\mbox{\includegraphics[height=0.26\textwidth,trim=13.5cm 0 30cm 0,clip]{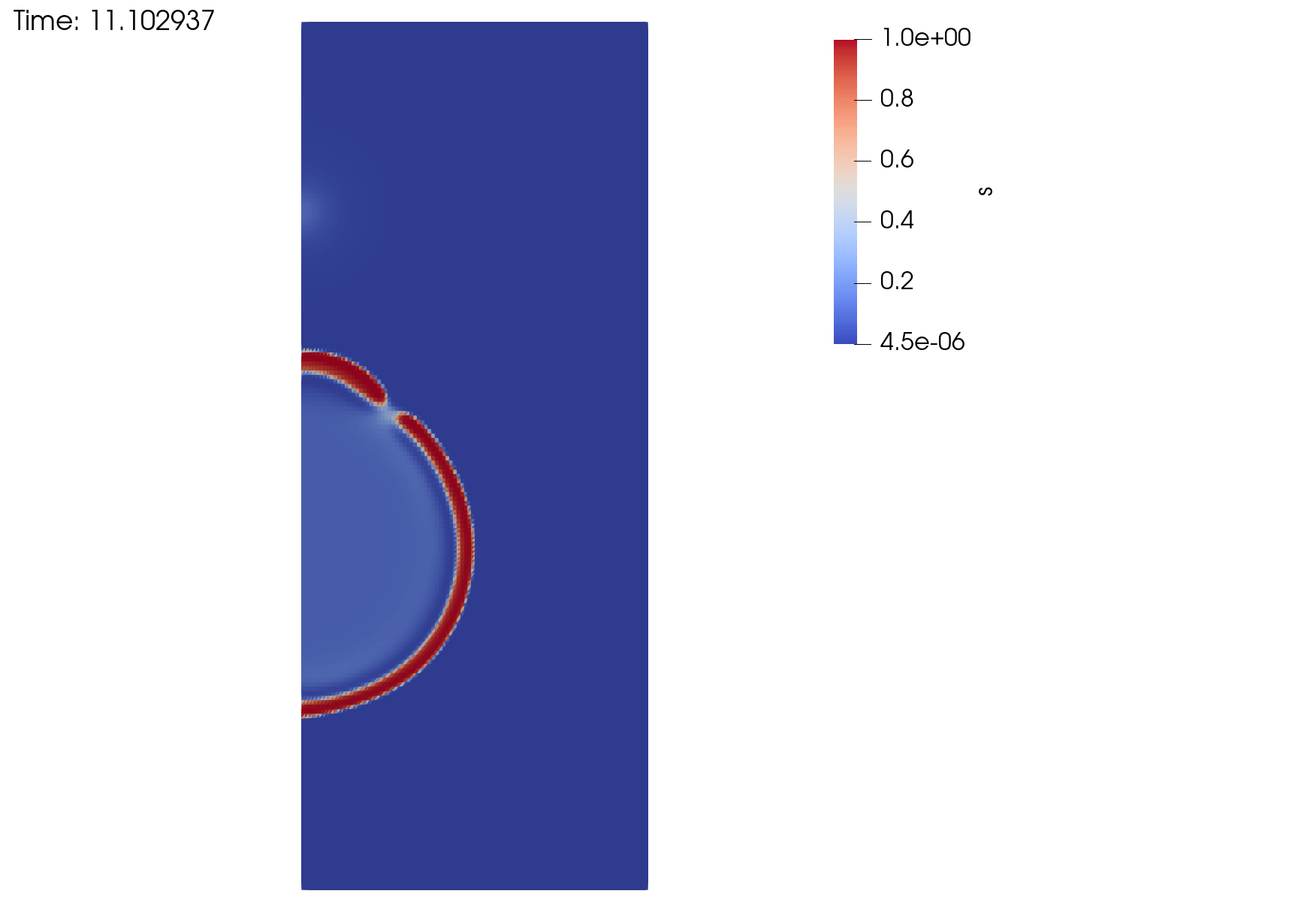}}\hfill%
\mbox{\includegraphics[height=0.26\textwidth,trim=13.5cm 0 20cm 0,clip]{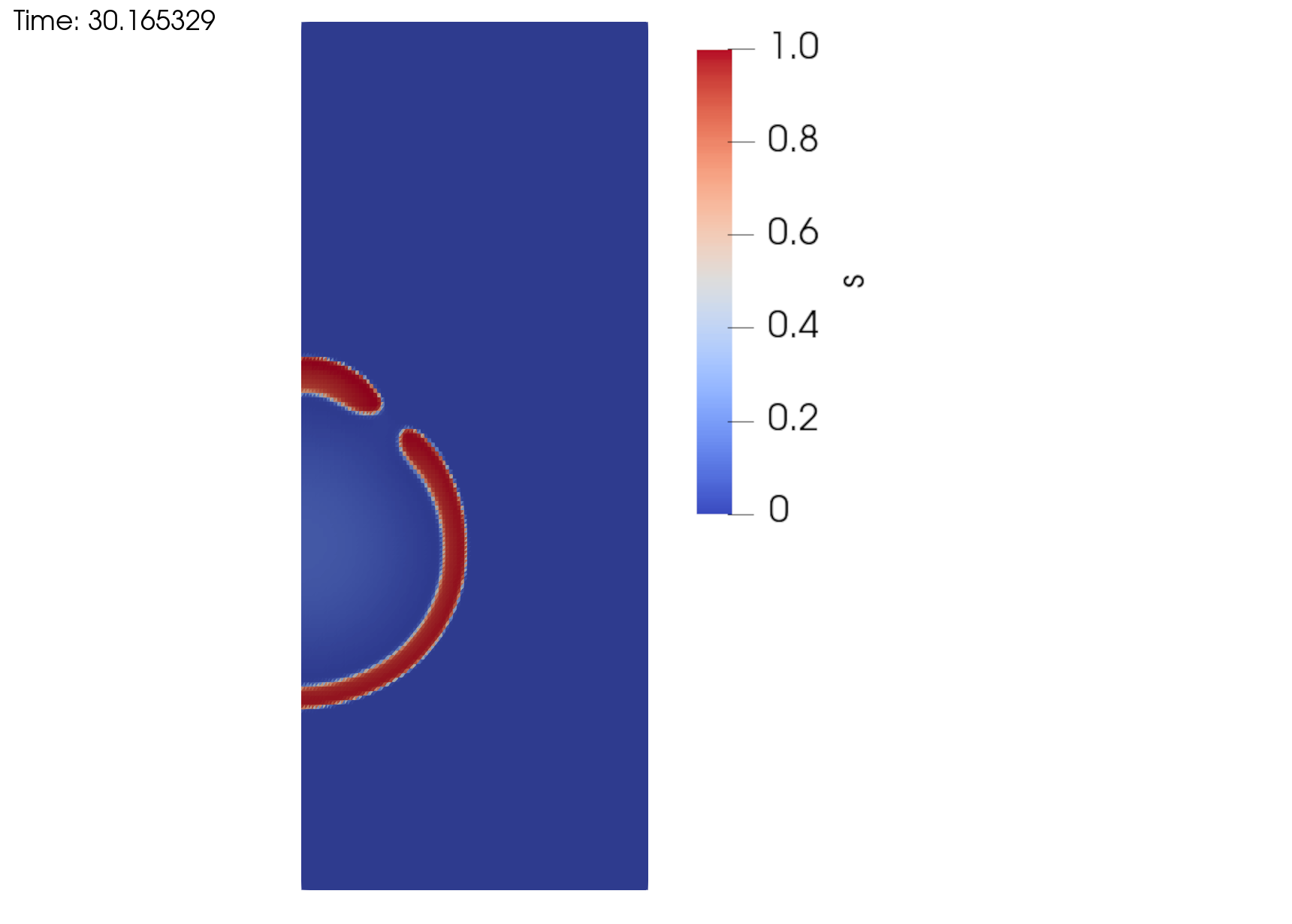}}
\hspace*{-1cm}
\begin{tikzpicture}[baseline=-1.3ex,>=stealth]
  \draw[->] (0,0) -- (0.35,0) node[right] {$r$};
  \draw[->] (0,0) -- (0,0.65) node[above] {$z$};
\end{tikzpicture}
\caption{Axisymmetric solute concentration corresponding to Fig.~\ref{fig:3ddroplet} with time increasing from left to right.}
\label{fig:3ddroplet_salty}
\end{figure}

We so far have considered spherically symmetric evaporating droplets, which constitutes  an idealized setting. In many realistic scenarios the symmetry is broken by interaction with other droplets, by external forces, or by spatial heterogeneity. In such cases, the topology of drying droplets can become considerably more complex and may strongly depend on the initial conditions. As an example of such a more complex scenario, in \Cref{fig:3ddroplet}, we show an evaporating solute-containing droplet that  merges with a much smaller droplet. The resulting symmetry breaking leads first to the formation of a crystal and later to the asymmetric development of a crystalline shell, which persists even at late stages of evaporation. This example illustrates that a full understanding of droplet evaporation requires the consideration of three-dimensional dynamics, in particular to capture  configurations that can arise from heterogeneous or multi-droplet initial states. As seen in \Cref{fig:3ddroplet}, the Cahn-Hilliard model captures the evolution of the vapor-liquid interface, including the pinch-off of a liquid bridge at $t=7.5$. Simultaneously, evaporation affects the solute concentration within the liquid phase through the coupled mass balance equation. Due to the asymmetric geometry, evaporation rates vary spatially, leading to non-uniform solute accumulation, with higher concentration near the liquid-vapor interface. Once the concentration exceeds the saturation threshold, crystallization becomes energetically favorable and the Allen-Cahn reaction kinetics are activated. At $t=9.1$, localized crystal regions (shown in red) appear and subsequently grow according to the state-dependent reaction rate. Since the crystalline phase is assumed to have zero mobility, it remains stationary and gradually forms a surface crust, which is visible at $t=11.1$ and appears thicker at $t=30.2$. The corresponding solute concentration in \Cref{fig:3ddroplet_salty} reveals the complex dynamics of an oversaturated evaporating droplet, whose numerical treatment and mathematical analysis become challenging due to the low regularity and strong spatial localization of the salt concentration field.


\begin{acknowledgments}
This research has been partially funded by Deutsche Forschungsgemeinschaft (DFG)
through grant CRC 1114 ``Scaling Cascades in Complex Systems'', Project Number
235221301, Project C02 ``Interface dynamics: Bridging stochastic and
hydrodynamic descriptions''. DP thanks for the funding within the DFG Priority Program SPP 2171 ``Dynamic Wetting of Flexible, Adaptive, and Switchable Surfaces'', Project Number 422792530. The authors would also like to acknowledge the anonymous referees for the helpful comments. 
\end{acknowledgments}

\bibliography{bibliography}

\end{document}